
%
\input amstex
\documentstyle{amsppt}
\magnification=1200
\vsize=21.5truecm
\hsize=16truecm
\parskip=6pt
%
%
%
%
\def\R{\Bbb R }
\def\N{\Bbb N }
\def\P{\Bbb P }
\def\Q{\Bbb Q }
\def\Z{\Bbb Z }
\def\C{\Bbb C }

\def\K{\Bbb K }
\def\F{\Bbb F }
%
%
%
%
\define\np{\vfil\eject}
\define\nl{\hfil\newline}
\redefine\l{\lambda}

\define\a{\alpha}

\redefine\b{\beta}

\define\x{\bold x}
\define\mapleft#1{\smash{\mathop{\longleftarrow}\limits^{#1}}}
\define\mapright#1{\smash{\mathop{\longrightarrow}\limits^{#1}}}
\define\mapdown#1{\Big\downarrow\rlap{
   $\vcenter{\hbox{$\scriptstyle#1$}}$}}
\define\mapup#1{\Big\uparrow\rlap{
   $\vcenter{\hbox{$\scriptstyle#1$}}$}}
\define\im{\text{Im\kern1.0pt }}
\define\re{\text{Re\kern1.0pt }}
\define\Kxn{\K[X_1,X_2,\ldots,X_n]}
\define\Kym{\K[Y_1,Y_2,\ldots,Y_m]}

\define\Cxy{\C[X,Y]}

\define\Mxn{(X_1-\a_1,\;X_2-\a_2,\;\ldots,\;X_n-\a_n)}
\define\Ox{{{\Cal O}_X}}

\define\Or{{{\Cal O}_R}}
\define\Os{{{\Cal O}_S}}
%
\def\Rad{\operatorname{Rad}}
\def\nil{\operatorname{nil}}
\def\Ann{\operatorname{Ann}}
\def\Spec{\operatorname{Spec}}
\def\Max{\operatorname{Max}}
\def\Priv{\operatorname{Priv}}
\def\Hom{\operatorname{Hom}}
\def\Quot{\operatorname{Quot}}
\def\Proj{\operatorname{Proj}}
%
%
\NoBlackBoxes
\TagsOnRight
\hfill Mannheimer Manuskripte 177

\hfill gk-mp-9403/3
\vskip 0.7cm
\topmatter
\title
Some Concepts of Modern Algebraic Geometry: Point, Ideal
and Homomorphism
\endtitle
\rightheadtext{Concepts of Modern Algebraic Geometry
}
\leftheadtext{Martin Schlichenmaier}
\author
Martin  Schlichenmaier\endauthor
\address
Martin Schlichenmaier,
Department of Mathematics and Computer Science,
University of Mannheim
D-68131 Mannheim, Germany
\endaddress
\email
schlichenmaier\@math.uni-mannheim.de
\endemail
\date May 94
\enddate
\endtopmatter
%
%
%
\document
%
%
%
%
\def\Invent{Invent.~Math.}

\def\NP{Nucl\. Phys\. B}

\def\Pnas{Proc\. Natl\. Acad\. Sci\. USA}
%
%
%
%
\def\refARAS{
\ref\key \ARAS\by Artin, M.
\book Algebraic Spaces
\bookinfo Yale Mathematical Monographs 3
\publaddr
New Haven, London
\publ Yale University Press\yr 1971
\endref}
\def\refART{
\ref\key \ART\by Artin, M.
\paper Geometry of quantum planes
\inbook
Azumaya algebras, actions, and modules
\bookinfo Proceedings Bloomington 1990, Contemp. Math. 124
\pages 1-15
\publaddr
Providence
\publ AMS
\endref}
\def\refBOGARE{
\ref\key \BOGARE\by Borho, W., Gabriel, P., Rentschler, R.
\book Primideale in Einh\"ullenden aufl\"osbarer Lie-Algebren
\bookinfo Lecture Notes in Mathematics 357
\publaddr
Berlin, Heidelberg, New York
\publ Springer\yr 1973
\endref}
\def\refEGAI{
\ref\key \EGAI\by  Grothendieck, A., Dieudonn\'e, J.~A.
\book  El\'ements de g\'eom\'etrie alg\'ebrique I
\publaddr
Berlin, Heidelberg, New York
\publ Springer\yr 1971
\endref}
\def\refEGAA{
\ref\key \EGAA\by  Grothendieck, A., Dieudonn\'e, J.~A.
\paper  El\'ements de g\'eom\'etrie alg\'ebrique
\jour Publi\-cations Math\'e\-ma\-tiques de l'Institute des Hautes
\'Etudes Scientifiques
\vol 8, 11, 17, 20, 24, 28, 32
\endref}
\def\refEHS{
\ref\key \EHS\by Eisenbud, D., Harris, J.
\book Schemes: The language of modern algebraic geometry
\publaddr
Pacific Grove, California
\publ Wadsworth \&Brooks\yr 1992
\endref}
\def\refGHPA{
\ref\key \GHPA\by Griffiths, Ph., Harris J.
\book Principles of algebraic geometry
\publaddr
New York
\publ John Wiley\yr 1978
\endref}
\def\refGOWA{
\ref\key \GOWA\by Goodearl, K.R., Warfield, R.B.
\book An introduction to noncommutative noetherian rings
\bookinfo London Math. Soc. Student Texts 16
\publaddr
New York, Port Chester, Melbourne, Sydney
\publ Cambridge University Press\yr 1989
\endref}
\def\refHAG{
\ref\key \HAG\by Hartshorne, R.
\book Algebraic geometry
\publaddr
Berlin, Heidelberg, New York
\publ Springer\yr 1977
\endref}
\def\refKNAS{
\ref\key \KNAS\by Knutson, D.
\book Algebraic Spaces
\bookinfo  Lecture Notes in Mathematics 203
\publaddr
Berlin, Heidelberg, New York
\publ Springer\yr 1971
\endref}
\def\refKUKA{
\ref\key \KUKA\by Kunz, E.
\book Einf\"uhrung in die kommutative Algebra
und algebraische Geometrie
\publaddr
Braunschweig
\publ Vieweg\yr 1979
\moreref
\book
Introduction to commutative algebra and algebraic geometry
\publaddr
Boston, Basel, Stuttgart
\publ Birkh\"auser\yr 1985
\endref}
\def\refMADR{
\ref\key \MADR\by Manin, Y. I.
\paper Notes on quantum groups and quantum de Rham complexes
\paperinfo MPI/91-60
\endref}
\def\refMANG{
\ref\key \MANG\by Manin, Y. I.
\book Topics in  noncommutative geometry
\bookinfo M. B. Porter Lectures
\publaddr
Princeton, New Jersey
\publ Princeton University Press\yr 1991
\endref}
\def\refMAQG{
\ref\key \MAQG\by Manin, Y. I.
\book Quantum groups and  noncommutative geometry
\publaddr
Montr\'eal
\publ Universit\'e de Montr\'eal (CRM)\yr 1988
\endref}
\def\refMRB{
\ref\key \MRB\by Mumford, D.
\book The red book of varieties and schemes
\bookinfo  Lecture Notes in Mathematics 1358
(reprint)\publaddr
Berlin, Heidelberg, New York
\publ Springer\yr 1988
\endref}

\def\refMPMP{
\ref\key \MPMP\by Mumford, D.
\paper Picard groups of moduli problems
\inbook Arithmetical algebraic geometry (Purdue 1963)
\pages 33-38
\ed O.~F.~G.,~Schilling
\publ Harper \&Row,
\publaddr New York\yr 1965
\endref}
\def\refROS {
\ref\key \ROS\by Rosenberg, Alexander, L.
\paper The left spectrum, the Levitzki radical, and
noncommutative schemes
\jour \Pnas\vol 87\issue 4\pages 8583-8586
\yr 1990
\endref}
\def\refSCHLRS{
\ref\key \SCHLRS\by Schlichenmaier,~M.
\book An introduction to Riemann surfaces, algebraic curves
and moduli spaces
\bookinfo  Lecture Notes in Physics 322
\publaddr
Berlin, Heidelberg, New York
\publ Springer\yr 1989
\endref}
\def\refVIST{
\ref\key \VIST\by Vistoli, A.
\paper Intersection theory on algebraic stacks and
their moduli spaces, Appendix
\jour \Invent
\vol 97\pages 613 -- 670 \yr 1989
\endref}
\def\refWEQG{
\ref\key \WEQG\by Wess, J., Zumino, B.
\paper Covariant differential calculus on the quantum hyperplane
\jour
\NP,\  Proceed\. Suppl\. \vol 18B\yr 1990\page 302
\endref}
%
\widestnumber\key{AAAAA}
\def\GHPA{GH} 
\def\EHS{EH}  
\def\KUKA{Ku} 
\def\MRB{Mu-1}   
\def\HAG{H}   
\def\EGAI{EGA I} 
\def\EGAA{EGA} 
\def\SCHLRS{Sch}  
\def\GOWA{GoWa}   
\def\MAQG{Ma-1}     
\def\MANG{Ma-2}     
\def\MADR{Ma-3}     
\def\MPMP{Mu-2}  
\def\ARAS{Ar-1}     
\def\VIST{Vi}     
\def\KNAS{Kn}     
\def\WEQG{WZ}     
\def\BOGARE{BGR}     
\def\ART{Ar-2}     
\def\ROS{R}        
%
%
\vskip 1cm
\head
{\bf Preface}
\endhead
\vskip 0.7cm
This is a write-up of lectures given at the ``Kleine
Herbstschule 93'' of the Graduier\-ten\-kolleg ``Mathematik im
Bereich Ihrer Wechselwirkungen  mit der  Physik''
at the  Ludwig-Maximilians-Universit\"at M\"unchen.
Starting from classical algebraic geometry over the complex numbers
 (as it can be found for
example in \cite{\GHPA})
it was the goal of these lectures to introduce some
concepts
of the modern point of view in algebraic geometry. Of course, it was quite
impossible even to give  an introduction to the whole subject
in such a limited time.
For this reason the lectures and now the write-up concentrate on
the substitution  of the concept of  classical points by the
notion of ideals  and homomorphisms of algebras.

These concepts were established by
Grothendieck in the 60s.
In the following they were proven to be very fruitful in mathematics.
I do not want to give an historic account of this claim. Let me just
mention the  proof of the
Weil conjectures by Pierre Deligne
(see \cite{\HAG,\rm App.C})
and the three more recent results:
Faltings' proof of Mordell's conjecture,
Faltings' proof of the Verlinde formula  and Wiles'
work in direction towards Fermat's Last Theorem.
\footnote{At the time this is written it is not clear whether the
gap found in Wiles' ``proof'' really can be closed.}
But also in theoretical physics, especially in connection with
the theory of quantum groups and noncommutative geometries,
it was necessary to extend the concept of points.
This is one reason for the
increasing interest in modern algebraic geometry
among theoretical
physicists.
Unfortunately, to enter the field is not an easy task. It has its own
very well developed language and tools. To enter it in a linear
way if it would be possible at all (which I doubt very much) would
take a prohibitive long time.
The aim of the lectures was to decrease the barriers
at least a little bit and to make some appetite for further studies
on a beautiful subject.
I am aiming at mathematicians and    theoretical physicists
who want to gain some feeling and some understanding of
these concepts. There is nothing new for algebraic geometers  here.

What are the prerequisites? I only assume some general basics of
mathematics (mani\-folds, complex variables, some algebra).
I try to stay elementary and hence assume only few facts from
algebraic geometry. All of these can be found in the first few chapters
of \cite{\SCHLRS}.

The write-up follows very closely the material presented at the
lectures. I did withstand the temptation to reorganize the material
to make it more systematic, to supply all proofs, and to add other
important topics.
Especially the infinitesimal and the global aspects are still missing.
Such an extension would considerably increase the amount of pages
and hence obscure the initial goal to give a short introduction
to the subject and to make appetite for further self-study.
What  made it easier for me to  decide in this way is
that there is a recent little book
by Eisenbud and Harris available now
\cite{\EHS} which  (at least that is what I hope) one should be able
to study with profit after these lectures.
The book \cite{\EHS} substitutes (at least partially) the for a
long period only available pedagogical introduction to the
language of schemes, the famous red book of varieties and
schemes by Mumford \cite{\MRB}.
\footnote{Which is still very much recommended to be read.
Recently, it has been reprinted in the Springer Lecture Notes Series.}
If you are looking for more details you can either consult Hartshorne
\cite{\HAG} or directly Grothendieck \cite{\EGAI},\cite{\EGAA}.
Of course, other good sources are available now.

Finally, let me thank the audience for their active listening and the
organizers of the Herbstschule for the invitation.
It is a pleasure for me to give special thanks to
Prof.~M.~Schottenloher and Prof.~J.~Wess.
\vskip 1cm
\head
{\bf 1. Varieties}
\endhead
\vskip 0.7cm
As we know from school the geometry of the plane consists of points, lines,
curves, etc\. with certain relations between them.
The introduction of coordinates (i.e\. numbers) to ``name'' the points
has been proven to be very useful.
In the real plane every point can be uniquely described by its
pair $(\a,\b)$ of Cartesian coordinates. Here $\a$ and $\b$ are
real numbers. Curves are ``certain'' subset of
$\ \R\times\R=\R^2$.
The notion ``certain'' is of course very unsatisfactory.

In classical algebraic geometry the subsets
defining the geometry are the set of
points where a given set of polynomials have a common
zero (if we plug in the coordinates of the points in the polynomial).
To give an example: the polynomials $X$ and $Y$ are elements of
the polynomial ring in 2 variables
over the real numbers $\R$.
They define the following polynomial functions:
$$X,\ Y:\ \R^2\to\R,\qquad
(\a,\b)\mapsto X(\a,\b)=\a,\ \text{\ resp.}\quad Y(\a,\b)=\b\ .$$
These two functions are called coordinate functions.
The point $(\a_0,\b_0)\in\R^2$ can  be given as zero set
$$\{(\a,\b)\in\R^2\mid X(\a,\b)-\a_0=0,\ Y(\a,\b)-\b_0=0\ \}\ .$$

Let me  come to the general definition. For this let
$\K$ be an arbitrary  field (e.g\. $\C,\;\R,\;\Q,\;\F_p,\;
\overline{\F_p},\ldots$)
and $\ \K^n=\underbrace{\K\times\K\times\cdots\times\K}_{
\text{$n$ times}}\ $ the $n-$dimensional {\sl affine
space} over $\K$.
I shall describe the objects of the geometry as zero sets of polynomials.
For this let $R_n=\K[X_1,X_2,\ldots,X_n]$ be the polynomial ring in
$n$ variables.
A subset $A$ of $\K^n$ should be a geometric object if there
exist finitely many polynomials $f_1,f_2,\ldots,f_s\in R_n$ such that
$$\x\in A\qquad\text{if and only if}\qquad
f_1(\x)=f_2(\x)=\cdots=f_s(\x)=0\ .$$
Here and in the following it is understood that
$\x=(x_1,x_2,\ldots,x_n)\in \K^n$ and $f(\x)\in\K$ denotes the
number obtained by replacing the variable $X_1$ by the number $x_1$, etc\..

Using the notion of ideals it is possible to define these sets $A$ in a
 more elegant fashion.
An ideal of an arbitrary ring $R$ is a subset of
$R$ which is closed
under addition :
$\ I+I\subseteq I\ $, and under multiplication with the whole ring:
$\ R\cdot I\subseteq I\ $.
A good reference to recall the necessary prerequisites from
algebra is \cite{\KUKA}.
Now let $I=(f_1,f_2,\ldots,f_r)$ be the ideal generated by the polynomials
$\ f_1,f_2,\ldots,f_s\  $
which define $A$, e.g\.
$$I=R\cdot f_1+R\cdot f_2+\cdots+R\cdot f_s=
\{r_1f_1+r_2f_2+\cdots+r_sf_s\mid r_i\in R, \ i=1,\ldots,s\}\ .$$
\definition{Definition}
A subset $A$ of $\ \K^n\ $ is called an {\sl algebraic set} if there is
an ideal $I$ of $R_n$ such that
$$\x\in A\iff f(\x)=0\quad\text{for all}\ f\in I .$$
The set $A$ is  called  the {\sl vanishing set of the ideal $I$},
in symbols $A=V(I)$ with
$$V(I):=\{\,\x\in
\K^n\mid f(\x)=0,\ \forall f\in I\;\}\ .\tag 1-1$$
\enddefinition
\remark{Remark 1}
It is enough to test the vanishing with respect to the generators
of the ideal in the definition.
\endremark
\remark{Remark 2}
There is no finiteness condition mentioned in the definition.
Indeed this is not necessary, because the polynomial ring
$R_n$
is a noetherian ring.
Recall a ring is a {\sl noetherian ring}  if every ideal has a
finite set of generators. There are other useful equivalent definitions
of a noetherian ring. Let me here recall only the fact that every
strictly ascending chain of ideals (starting from one
ideal)  consists only of finitely many ideals.
But every field $\K$ has only the (trivial) ideals
$\{0\}$ and $\K$ (why?), hence $\K$ is noetherian.
Trivially, all principal ideal rings (i.e\. rings where every ideal can be
generated by just one element) are noetherian.
Beside the fields there are two important examples of principal
ideal rings: $\Z$ the integers, and $\K[X]$ the polynomial ring in
one variable over the field $\K$. Let me recall the proof for $\Z$.
Take  $I$  an ideal of $\Z$. If $I=\{0\}$ we are done. Hence assume
$I\ne \{0\}$ then there is a $n\in\N$ with $n\in I$ minimal.
We now claim $I=(n)$. To see this take $m\in I$. By the  division
algorithm of Euklid there are $q,r\in\Z$ with $0\le r<n$ such that
$\ m=qn+r\ $. Hence, with $m$ and $n$ in $I$ we get $r=m-qn\in I$.
 But $n$ was chosen
minimal, hence $r=0$ and $m\in (n)$.
Note that the proof for $\K[X]$ is completely analogous if we
replace the division algorithm for the integers by the
division algorithm for polynomials.

Now we have
\proclaim{Hilbertscher Basissatz}
Let $R$ be a noetherian Ring. Then $R[X]$ is also noetherian.
\endproclaim
\noindent
As a nice exercise you may try to proof it by yourself
(maybe guided by \cite{\KUKA}).
\endremark
\remark{Remark 3}
If $R$ is a noncommutative ring  one has to deal with left, right and
two-sided ideals. It is also necessary to define left, right, and
two-sided noetherian.
\endremark

It is time to  give  some {\bf examples of algebraic sets}:

\noindent
({\bf 1}) The whole affine space is the zero set of
the zero ideal: $\K^n=V({0})$.

\noindent
({\bf 2}) The empty set is the zero set of the whole ring $R_n$:
$\emptyset=V((1))$.

\noindent
({\bf 3}) Let $\a=(\a_1,\a_2,\ldots,\a_n)\in\K^n$ be a point given
by its coordinates. Define the ideal
$$I_\a=(X_1-\a_1,X_2-\a_2,\ldots,X_n-\a_n),$$ then
$\{\a\}=V(I_\a)$.

\noindent
({\bf 4}) Now take 2 points $\a,\b$ and their associated ideals $I_\a,I_\b$
 as defined in (3). Then $\ I_\a\cap I_\b\ $ is again an ideal
and we get $\ \{\a,\b\}=V(I_\a\cap I_\b)\ $.
\nl
This is a {\bf general fact}. Let $A=V(I)$ and $B=V(J)$ be two algebraic
sets then the union $A\cup B$ is again an algebraic set because
$\ A\cup B=V(I\cap J)$. Let me give a proof of this.
Obviously, we get for two ideals $K$ and $L$ with
$\ K\subseteq L\ $ for their vanishing sets $V(K)\supseteq V(L)$. Hence
because $\ I\cap J\subseteq I\ $ and $\ I\cap J\subseteq J\ $
we obtain $\ V(I\cap J)\supseteq V(I)\cup V(J)$. To proof the other
inclusion assume that there is an $x\not\in  V(I)\cup V(J)$ then there are
$f\in I$ and $g\in J$ with $f(x)\ne 0$ and $g(x)\ne 0$. Now $f\cdot
g\in I\cap J$ but $(f\cdot g)(x)=f(x)\cdot g(x)\ne 0$. Hence
 $x\notin V(I\cap J)$.
Let me repeat the result for further reference:
$$V(I)\cup V(J)\quad=\quad V(I\cap J)\ .\tag 1-2$$

({\bf 5}) A hypersurface $H$ is  the vanishing set of the ideal generated
by a single polynomial $f$: $H=V((f))$.
An example in $\C^2$ is given by $I=(Y^2-4X^3+g_2X+g_3)$
where $g_2,g_3\in\C$. The set
$V(I)$ defines a cubic curve in the plane. For general $g_2,g_3$
this curve is isomorphic to a (complex)
 one-dimensional torus with the point $0$ removed.

({\bf 6}) Linear affine subspaces
 are algebraic sets. A linear affine subspace  of $\K^n$
is the set of   solutions of a system of linear equations
$\ A\cdot \x=\bold b\ $ with
$$A=\pmatrix \bold a_{1,*}\\ \cdots \\ \bold a_{r,*}
\endpmatrix,\quad b=\pmatrix b_1\\ \cdots \\ b_r
\endpmatrix,\quad
\bold a_{i,*}\in \K^n,\quad b_i\in \K\ ,i=1,\ldots,r\ .$$
The solutions (by definition) are given as the elements of
the vanishing set of the ideal
$$I=(\bold a_{1,*}\cdot X-b_1,\ \bold a_{2,*}\cdot X-b_2,\
\cdots,\  \bold a_{r,*}\cdot X-b_r)\ .$$

({\bf 7}) A special case are the straight lines in the plane.
For this let $l_i=a_{i,1}X+a_{i,2}Y-b_i$,  $i=1,2$ be two
linear forms. Then $L_i=V((l_i))$, $i=1,2$ are lines.
For the union of the two lines we obtain by (1-2)
$$L_1\cup L_2=V((l_1)\cap (l_2))=V((l_1\cdot l_2))\ .$$
Note that  I do not claim  $(l_1)\cap (l_2)=(l_1\cdot l_2)$.
The reader is encouraged to search for conditions when this
will hold.
For the intersection of the two lines
we get  $L_1\cap L_2=V((l_1,l_2))$ which can be written as
$V((l_1)+(l_2))$. Of course, this set consists just of one point
if the linear forms $l_1$ and $l_2$ are linearly independent.
Again, there is the {\bf general fact}
$$V(I)\cap V(J)=V(I+J),\tag 1-3$$
where
$$I+J:=\{\,f+g\mid f\in I,g\in J\,\}\ .$$
\bigskip
You see there is a ample supply of examples for algebraic sets.
Now we introduce for $\K^n$ a topology, the {\sl Zariski-Topology}.
For this we call a subset $U$ open if it is a complement
of an algebraic set, i.e\. $\ U=\K^n\setminus V(I)\ $ where
$I$ is an ideal of $R_n$. In other words: the closed sets are the
algebraic sets.
It is easy to verify the axioms for a topology:

\roster
\item
$\K^n$ and $\emptyset$ are open.
\item
Finite intersections are open:
$$U_1\cap U_2=(\K^n\setminus V(I_1))\cap (\K^n\setminus V(I_2))
=\K^n\setminus (V(I_1)\cup V(I_2) )=  \K^n\setminus V(I_1\cap I_2)\ .$$
\item
Arbitrary unions are open:
$$\bigcup_{i\in S}(\K^n\setminus V(I_i))=\K^n\setminus\bigcap_{i\in S}V(I_i)=
\K^n\setminus V(\sum_{i\in S}I_i) \ .$$
\endroster

Here $S$ is allowed to be an infinite index set. The ideal
$\ \sum_{i\in S}I_i\ $ consists of elements in $R_n$ which are finite
sums of elements belonging to different $I_i$.
The claim (1-3) easily extends to this setting.
\bigskip
Let us study the affine line $\K$. Here $R_1=\K[X]$. All ideals in
$K[X]$ are principal ideals, i.e\.  generated by just one
polynomial. The vanishing set of an ideal consists just of the finitely many
zeros of this polynomial (if it is not identically zero).
Conversely, for every set of finitely many points there is
a polynomial vanishing exactly at these points.
Hence, beside the empty-set and the whole line
the algebraic sets are the
sets of finitely many points.
At this level there is already a new concept showing up.
The polynomial assigned to a certain point set is not unique.
For example it is possible to increase the vanishing order of the polynomial
at a certain zero without changing the vanishing set.
It would  be better to talk about point sets with
multiplicities to
get a closer correspondence to the polynomials.
Additionally, if $\K$ is not algebraically closed then there are
non-trivial polynomials without any zero at all.
 These ideas we will
take up in later lectures.
The other important observation is that the open sets in $\K$ are
either empty or dense. The latter says that the closure $\overline{U}$
of $U$, i.e\. the smallest closed set which contains $U$ is
the whole space $\K$.
Assuming the
whole space to be irreducible this is  true in a more general context.
\definition{Definition}\nl
(a) Let $V$ be a closed set. $V$ is called {\sl irreducible}
if for every decomposition
$\  V=V_1\cup V_2\ $ with $V_1,V_2$ closed we have
$ V_1=V\ $ or $\ V_2=V\ $.\nl
(b) An algebraic set which is irreducible is called a {\sl variety}.
\enddefinition
Now let $U$ be an open subset of  an irreducible $V$. The two set
$\ V\setminus U\ $ and $\ \overline{U}\ $ are closed and $V=(V\setminus U)\cup
\overline{U}$. Hence, $V$ has to be one of these sets.
Hence,
either $U=\emptyset$ or $V=\overline{U}$. As promised,
this shows that every open subset of an irreducible space is either
empty or dense.
Note that this has nothing to do with our special situation. It follows
from general topological arguments.
In the next section we will see that the spaces $\K^n$ are
irreducible.
\bigskip
Up to now we were able to describe our geometric objects
with the help of the ring of polynomials. This ring
plays  another important role in the whole theory.
We need it
to study polynomial (algebraic) functions on $\K^n$.
If $f\in R_n$ is a polynomial then
$\  \x\mapsto f(\x)\ $ defines a map from $\K^n$ to $\K$.
This can be extended to functions on algebraic sets
$A=V(I)$.
We associate to $A$ the quotient ring
$$R(A):=\K[X_1,X_2,\ldots,X_n]/I\ .$$
This ring is called the coordinate ring of $A$.
The elements of $R(A)$ can be considered as functions
on $A$. Take
$\x\in A$, and $\bar f\in R(A)$ then $\bar f(\x):=f(\x)$
is a
well-defined element of $\K$. Assume $\bar f=\bar g$ then there is an $h\in I$
with $f=g+h$ hence $f(\x)=g(\x)+h(\x)=g(\x)+0$.
You might have noticed that it is not really correct to
call this ring the coordinate ring of $A$.
It is not clear, in fact it is not even true that the ideal $I$ is fixed
by the set $A$. But $\ R(A)\ $ depends on $I$.
A first way
to avoid these complications is to assign to every $A$ a unique
defining ideal,
$$I(A):=\{f\in R_n\mid f(\x)=0,\forall \x\in A\}\ .\tag 1-4$$
It is the largest ideal which defines $A$.
For arbitrary ideals we always obtain  $I(V(I))\supseteq I$.

There is a second possibility which even takes advantage out of
the non-uniqueness.
We could have added the additional data of the defining ideal $I$
in the notation.
Just simply assume that when we use $A$ it comes with
a certain $I$.
 Compare this with the situation above where
we determined the closed sets of $\K$.
Again this  at the first glance
annoying fact of non-uniqueness of $I$ will allow us to introduce
multiplicities in the following which in turn will be rather useful
as we will see.

Here another warning is in order. The elements of $\ R(A)\ $ define
usual functions on the set $A$. But different elements can define
the same function. In particular, $R(A)$ can have zero divisors and
nilpotent elements (which always give the zero function).

The ring $R(A)$ contains all the geometry of $A$.
As an example, take  $\ A\ $ to be a curve in the plane and
$\ P\ $ a point in the plane.
Then $A=V((f))$ with $f$ a polynomial in $X$ and $Y$ and
$\ P=V((X-\a,Y-\b))\ $. Now $\ P\subset A\ $
(which says that the point $P$ lies on $A$) if and only if
$(X-\a,Y-\b)\supset (f)$. Moreover,
in this case  we obtain the following
diagram of ring homomorphisms
$$
\CD
R_2 @> / (f)>> R(A) \\
@A \subseteq AA @A \subseteq AA \\
(X-\a,Y-\b)\quad @> / (f)>>\quad (X-\a,Y-\b)/(f) \\
@A \subseteq AA @A \subseteq AA \\
(f) @>>> \{0\}\ .
\endCD
$$
The quotient $(X-\a,Y-\b)/(f)$ is an ideal of $R(A)$ and corresponds
to the point $P$ lying on $A$.

Indeed, this is the general situation which we will study in the following
sections: the algebraic sets on $A$ correspond to the ideals of
$R(A)$ which in turn correspond to the ideals lying between the
defining ideal of $A$ and the whole ring $R_n$.

Let me close this section by studying the geometry of a single
point $P=(\a,\b)\in \K^2$. A defining ideal is
$I=(X-\a,Y-\b)$. If we require "multiplicity one" this is
the defining ideal.
Hence, the coordinate ring $R(P)$ of a point
is $\ \K[X,Y]/I\cong \K\ $.
The isomorphismus is induced  by the homomorphism
$\K[X,Y]\to\K$
given by $X\to \a,\ Y\to \b$.
Indeed, every element $r$ of $\ \K[X,Y]\ $
can be given as
$$r=r_0+(X-\a)\cdot g+(Y-\b)\cdot f,\quad r_0\in \K,\ f,g\in \K[X,Y]\
.\tag1-5$$
Under the homomorphism
$\ r\ $ maps to $\ r_0\ $.
Hence $\ r\ $ is in the kernel of the map
if and only if $r_0$  equals
$0$ which in turn is the case if and only if $r$ is in the ideal $I$.
The description (1-5) also shows  that $I$ is a maximal ideal.
We call an ideal $I$ a maximal ideal if there are no ideals
between $I$ and the whole ring $R$ (and $I\ne R$).
Any ideal strictly larger than
the above $I$ would contain an $r$ with $r_0\ne 0$.
Now this ideal would contain $r,(X-\a),(Y-\b)$ hence also $r_0$.
Hence also $(r_0)^{-1}\cdot r_0=1$. But an ideal containing $1$ is
always the whole ring.

On the geometric side the points are the minimal sets.
 On the level of the ideals in
$R_n$ this corresponds to the fact that an ideal
defining a point (with multiplicity one) is a maximal
ideal. If the field $\K$ is algebraically closed then every maximal
ideal corresponds indeed to a point.
\np
\vskip 1cm
\head
{\bf 2. The spectrum of a ring}
\endhead
\vskip 0.7cm
In the last lecture we saw that geometric objects are in
correspondence to algebraic objects of the coordinate ring.
This we will develop more systematically in this lecture.
We had the following correspondences (1-1), (1-4)
$$\align
\text{ideals of}\  R_n\quad&\mapright{V}\quad\text{algebraic sets}\\
\hbox{  }\\
\text{ideals of}\  R_n\quad&\mapleft{I}\quad\text{algebraic sets.}
\endalign
$$
Recall the definitions: ($R_n=\Kxn$)
$$
V(I):=\{\,\x\in \K^n\mid f(\x)=0,\ \forall f\in I\,\},\qquad
I(A):=\{\,f\in R_n\mid f(\x)=0,\forall \bold x\in A\,\}\ .
$$
In general $\ I(V(I))\ $ will be bigger than the ideal $I$.
Let me give an example.
Consider in $\C[X]$ the ideals $I_1=(X)$ and $I_2=(X^2)$.
Then $V(I_1)=V(I_2)=\{0\}$. Hence both ideals define the same
point as vanishing set.
Moreover $I(V(I_2))=I_1$ because $I_1$ is a maximal ideal.
If we write down the coordinate ring of the two situations
we obtain for $I_1$ the ring
$\C[X]/(X)\cong \C$. This is  the expected situation because the functions
on a point are just the constants.
For $I_2$ we obtain
$\ \C[X]/(X^2)\cong \C\oplus \C\cdot\epsilon\  $ the algebra
generated by $1$ and $\epsilon$ with the relation $\epsilon^2=0$
($X$ maps to $\epsilon$).
Hence, there is no 1-1 correspondence between ideals and algebraic sets.
If one wants such a correspondence one has to throw away the
"wrong" ideals. This is in fact possible (by considering the so called
radical ideals, see the definition below).
 Indeed, it is rather useful to allow all ideals
to obtain more general objects (which are very useful)
 as the classical objects.

To give an example:
take the affine real line and
let $I_t=(X^2-t^2)$ for $t\in\R$ be a family of ideals. The role of $t$ is
the role of a parameter one is allowed to vary.
Obviously,
$$I_t=((X-t)(X+t))=(X-t)\cdot(X+t).$$
For $t\ne 0$ we obtain  $V(I_t)=\{t,-t\}$ and for $t=0$ we obtain
$V(I_0)=\{0\}$. We see that for general values of $t$ we get
two points, and for  the
value $t=0$ one point. If we approach with $t$ the
value $0$
the two different points $\pm t$ come closer and closer together.
Now  our intuition says that the limit point $t=0$ better
should be counted twice.
This intuition we can make mathematically precise on the level of
the coordinate rings. Here we have
$$R_t=\R[X]/I_t\cong \R\oplus \R\cdot \epsilon,\quad
\epsilon^2=t^2\ .$$
The coordinate ring is a two-dimensional vector space over $\R$
which reflects the fact  that
we deal with two points. Everything here is also true for
the exceptional value $t=0$. Especially $R_0$ is again two-dimensional.
This says we  count the point $\{0\}$ twice.
The drawback is that the interpretation of the elements of
$\ R_t\ $ as classical functions will not be possible in all cases.
In our example for $t=0$ the element $\bar X$ will be nonzero
but $\bar X(0)=0$.
\bigskip
\noindent
{\it For the following definitions let $R$ be an
 arbitrary commutative ring with
unit 1.}
\medskip
\definition{Definition}

\noindent
(a) An ideal $P$ of $R$ is called a {\sl prime ideal} if $P\ne R$ and
$\ a\cdot b\in P\ $ implies $\ a\in P\ $ or $\ b\in P$.

\noindent
(b) An ideal $M$ of $R$ is called a {\sl maximal ideal} if $M\ne R$ and
for every ideal $M'$ with $M'\supseteq M$
it follows that $M'=M$ or $M'=R$.

\noindent
(c) Let $I$ be an ideal. The {\sl radical} of $I$ is defined as
$$\Rad(I):=\{\,f\in R\mid \exists n\in\N:f^n\in I\,\}\ .$$

\noindent
(d) The {\sl nil radical } of the ring $R$ is defined as
$\ \nil(R):=\Rad(\{0\})$ .

\noindent
(e) A ring is called {\sl reduced} if $\nil(R)=\{0\}$.

\noindent
(f) An ideal $I$ is called a {\sl radical ideal} if
$\ \Rad(I)=I$.
\enddefinition
\noindent
Starting from these definitions
there are a lot of easy exercises for the reader:

\noindent
(1) Let $P$ be a prime ideal. Show: $R/P$ is a ring without
zero divisor (such  rings are called integral domains).

\noindent
(2) Let $M$ be a maximal ideal. Show $R/M$ is a field.

\noindent
(3) Every  maximal ideal is a prime ideal.

\noindent
(4) $\Rad(I)$ is an ideal.

\noindent
(5)
$\ \Rad(I)\ $
equals the intersection of all prime ideals containing $I$.

\noindent
(6)  $\nil(R/I)=\Rad(I)/I$ and conclude that every prime ideal is
a radical ideal.

\noindent
(7) $\Rad I$ is a radical ideal.

\bigskip
Let me return to the rings $R_t$ defined above.
The ideals  $I_t$ are not  prime  because
neither $\ X+t\ $ nor $\ X-t\ $ are in $I_t$ but $\ (X+t)(X-t)\in I_t$.
In particular, $R_t$ is not an integral domain: $(\epsilon+t)(\epsilon-t)=0$.
Let us calculate $\nil(R_t)$.
For this we take an element $0\ne z=a+b\epsilon$ and calculate
$$0=(a+b\epsilon)^n=a^n+\binom n1 a^{n-1}b^1\epsilon+
\binom n2 a^{n-2}b^2\epsilon^2+\cdots \ .$$
Replacing $\epsilon^2$ by the positive real number $t^2$ we obtain
$$\gather
0=(a+b\epsilon)^n=\left(a^n+\binom n2 a^{n-2}b^2t^2+
\binom n4 a^{n-4}b^4t^4+\cdots\right) +
\\+\epsilon
 \left(\binom n1 a^{n-1}b^1+
\binom n3 a^{n-3}b^3t^2+\cdots\right)\ .
\endgather$$
{}From this we conclude that all terms in the first and in the second sum
 have to vanish (all terms have the same sign).
This implies $a=0$.
Regarding the last element in both sums
we see that for $t\ne 0$ we get $b=0$. Hence $\nil(R_t)=\{0\}$,
for $\ t\ne 0$
and the ring $R_t$ is reduced.
 For $t=0$ the value of $b$ is arbitrary.
 Hence $\nil(R_0)=(\epsilon)$, which says that  $R_0$ is
 not a
reduced ring. This is the typical situation:
a non-reduced coordinate ring $R(V)$ corresponds to a variety $V$ which
should be considered with higher multiplicity.
\bigskip
For the polynomial ring we have the following  very important result.
\proclaim{Hilbertscher Nullstellensatz}
Let $I$ be an ideal in $R_n=\Kxn$. If $\K$ is algebraically closed then
$I(V(I))=\Rad(I)$.
\endproclaim
\noindent
The proof of this theorem is not easy. The main tool is the
following version of the Nullstellensatz which more resembles
his name
\proclaim{Hilbertscher Nullstellensatz}
Let $I$ be an ideal in $R_n=\Kxn$.
$I\ne R_n$. If $\K$ is algebraically closed then
$\ V(I)\ne\emptyset$.
In other words given a set of polynomials such that the
constant polynomial 1 cannot be represented as a
$R_n-$linear sum in these polynomials then there is a simultaneous zero of
these polynomials.
\endproclaim
\noindent
For the proof let me refer to \cite{\KUKA}.

The Nullstellensatz gives us a correspondence between algebraic sets
in $\K^n$  and
the radical ideals of $R_n=\Kxn$.
If we consider the prime ideals we get
\proclaim{Proposition}
Let $P$ be a radical ideal.
Then $P$ is a prime ideal if and only if $V(P)$ is a variety.
\endproclaim
\noindent
Before we come to the proof of the proposition let me state the following
simple observation. For arbitrary subsets $S$ and $T$ of $\K^n$  the ideals
$I(S)$ and $I(T)$ can be defined completely in the same way
as in (1-4), i.e\.,
$$
I(S):=\{f\in R_n\mid f(\x)=0,\forall \x\in S\}\ .\tag 2-1
$$
It is easy to show that
$$
I(S\cup T)=I(S)\cap I(T),\quad\text{and}\quad
V(I(S))=\overline{S}\ .\tag 2-2$$
Here $\overline{S}$ denotes the topological closure of $S$, which is
the smallest (Zariski-)closed subset of $\K^n$ containing $S$.
\demo{Proof of the above proposition}
Let $P$ be a prime ideal and set $\ Y=V(P)\ $ then
$I(V(P))=\Rad(P)=P$ by the Nullstellensatz. Assuming $Y=Y_1\cup Y_2$
a closed decomposition of $Y$ then
$\ I(Y)=I(Y_1\cup Y_2)=I(Y_1)\cap I(Y_2)=P\ $. Because $P$ is prime
either $P=I(Y_1)$ or $P=I(Y_2)$. Assume the first then
$\ Y_1=V(I(Y_1))=V(P)=Y$ (using that $Y_1$ is closed).
\nl
Conversely: let $Y=V(P)$ be irreducible  with $P$ a radical ideal.
By the Nullstellensatz $P=\Rad(P)=I(Y)$.
Let $\ f\cdot g\in P$ then
$\ f\cdot g\ $ vanishes on $Y$. We can decompose
$\ Y=(Y\cap V(f))\cup (Y\cap V(g))\ $ into closed subset of $Y$.
By the irreducibility it has to coincide with one of them. Assume
with the first. But this implies that $V(f)\supseteq Y$ and hence
$f$ is identically zero on $Y$. We get $f\in I(Y)=P$.
This shows that $P$ is a prime ideal.
\qed\enddemo
Note the fact that we restricted the situation to radical
ideals corresponds to the fact that
varieties as sets have always multiplicity 1, hence they are always
``reduced''. To incorporate all ideals and hence ``nonreduced structures''
 we have
to use the language of schemes (see below).

Let us look at  the maximal ideals of $R_n=\Kxn$. (Still $\K$ is
assumed  to be  algebraically closed).
The same argument as in the two-dimensional case shows that
the ideals
$$M_\a=\Mxn$$
are maximal and that $R_n/M_\a\cong \K$. This is even
true if the field $\K$ is not algebraically closed.
Now let $M'$ be a maximal ideal. By the Nullstellensatz
(here algebraically closedness is important)
there is a common
zero $\a$ for all elements $f\in M'$.
Take $f\not\in M'$ then $R_n=(f,M')$. Now
$f(\a)=0$ would imply that $\a$ is a zero of all polynomials in $R_n$
which is impossible. Hence, every polynomial $f$ which vanishes at $\a$
lies in $M'$.
All elements in $M_\a$ have  $\a$ as a zero. This implies
$\ M_\a\subseteq M'\subsetneqq R_n\ $. By the maximality of $M_\a$ we
conclude $M_\a=M'$.

Everything can be generalized to an arbitrary variety  $A$
 over an algebraically closed field.
The points of $A$ correspond to the maximal ideals of $R_n$ lying
above the defining prime ideal $P$ of $A$.
They correspond exactly to the maximal ideals in $R(A)$.
All of them
can be given as $M_\a/P$. This can be extended to the varieties
of $\K^n$ lying on $A$. They correspond to the prime ideals
of $R_n$ lying between
the prime ideal $P$ and the whole ring.
They in turn can be identified with the prime ideals of $R(A)$.
\np
Coming back to
arbitrary rings  it is now quite useful to talk about dimensions.
\definition{Definition}
Let $R$ be a ring. The {\sl (Krull-) dimension} $\ \dim R\ $ of a ring
$R$ is defined as the maximal  length $r$ of all  strict
chains of prime ideals $P_i$ in $R$
$$P_0\ \subsetneqq \ P_1\  \subsetneqq \ P_2\
\ldots\
\subsetneqq\  P_r\ \subsetneqq \ R\ .$$
\enddefinition
\remark{Example 1}
For a field $\K$  the only (prime) ideals are
$\{0\}\subset \K$. Hence $\dim \K=0$.
\endremark
\remark{Example 2}
The dimension of $R_n= \Kxn=R(\K^n)$ is  $n$. This result one should
 expect
from a reasonable definition of dimension.
Indeed we have the chain of prime ideals
$$(0)\ \subsetneqq\  (X_1)\  \subsetneqq\  (X_1,X_2)\ \subsetneqq
\ \cdots\  \subsetneqq\  \Mxn\ \subsetneqq\  R_n \ .$$
Hence $\dim R_n\ge n$. With some more commutative algebra it
is possible to show the equality, see \cite{\KUKA,S.54}.
\endremark
\remark{Example 3}
As a special case one obtains $\dim \K[X]=1$.
Here the reason is a quite general result. Recall that
  $\K[X]$ is a principal
ideal ring without zero divisors.
 Hence, every ideal $I$ can be generated by one element
$f$. Assume $I$ to be a prime ideal, $I\ne \{0\}$ and let $M=(g)$  be
 a maximal ideal
lying above $I$.
We show that $I$ is already maximal.
Because $(f)\subseteq (g)$ we get $f=r\cdot g$.   But $I$ is
prime. This implies either $r$ or $g$ lies in $I$. If $g\in I$ we are done.
If $r\in I$ then $r=s\cdot f$ and $f=f\cdot s\cdot g$.
In a ring without zero divisor one is allowed to cancel common factors.
We obtain $1=s\cdot g$. Hence, $1\in M$ which contradicts the fact that $M$
is not allowed to be the whole ring.
{}From this it follows that $\dim \K[X]=1$.
Note that we did not make any reference to the special nature of the
polynomial ring here.

What are the conditions on $f$ assuring that the ideal $(f)$ is prime.
The necessary and sufficient condition is that
$f$ is irreducible but not a unit. This says if there is
decomposition $\ f=g\cdot h\ $ then either $g$ or $h$ has to be a unit
(i.e\. to be invertible)
which in our situation says that $g$ or $h$ must be a constant.
This can be seen in the following way. From the decomposition it
follows (using $(f)$ is prime) that either
$g$ or $h$ has to be in $(f)$ hence is a multiple of $f$.
By considering the degree we see that  the complementary factor
has degree zero and hence is a constant.
\nl
Conversely, let $f$ be irreducible but not a unit. Assume $g\cdot h\in(f)$,
then $\ g\cdot h=f\cdot r$.
In the polynomial ring we have unique factorization (up to units)
into irreducible elements. Hence, the factor $f$ is contained either in
$g$ or $h$. This shows the claim.
\endremark
\remark{Example 4}
The ring of integers $\Z$ is also a principal ideal ring without
zero divisor.
Again we obtain $\dim \Z=1$. In fact, the integers behave very much
(at least from the
point of view of algebraic geometry) like the affine line over a field.
What are the ``points'' of $\Z$? As already said the points should
correspond to the maximal ideals. Every prime ideal in $\Z$ is maximal.
An ideal $(n)$ is prime exactly if $n$ is a prime number.
Hence, the ``points'' of $\Z$ are the prime numbers.
\endremark
\bigskip
Now we want to introduce the Zariski topology on the set of all prime
ideals of a ring.
First we introduce the sets
$$
\align
\Spec( R)\ &:=\ \{\; P\mid P\text{ is a prime ideal of } R\;\},\\
\Max ( R)\ &:=\ \{\; P\mid P\text{ is a maximal ideal of } R\;\}\ .
\endalign
$$
The set $\Spec(R)$ contains in some sense all irreducible ``subvarieties''
of the ``geometric model'' of $R$.
Let $S$ be an arbitrary subset of $R$. We define
the associated subset of $\Spec( R)$
as the set consisting of the  prime ideals which contain $S$:
$$V(S)\ :=\ \{\;P\in \Spec( R)\mid P\supseteq S\,\}\ .\tag 2-3$$
The subsets of $\Spec(R)$ obtained in this way
are called the closed subsets.
It is obvious that $\ S\subseteq T\ $ implies $\ V(S)\supseteq V(T)\ $.
Clearly, $V(S)$ depends only of the ideal generated by $S$: $V((S))=V(S)$.

This defines a topology on $\Spec(R)$ the
{\sl Zariski topology}.

\noindent
(1) The whole space and the empty set are closed:
$V(0)=\Spec(R)$ and $V(1)=\emptyset$.

\noindent
(2) Arbitrary intersections of closed sets are again closed:
$$ \bigcap_{i\in J}V(S_i)=V(\bigcup_{i\in J}S)\ .\tag 2-4$$
\noindent
(3) Finite unions of closed set are  again closed:
$$ V(S_1)\cup V(S_2)=V((S_1)\cap (S_2))\ .\tag 2-5$$
\noindent
Let me just show (2-5) here.
Because $\ (S_1),(S_2)\supseteq (S_1)\cap (S_2)\ $ we get
$\ V(S_1)\cup V(S_2)\subseteq V((S_1)\cap (S_2))$.
Take $P\in V((S_1)\cap (S_2))$.This says
$P\supseteq (S_1)\cap (S_2)$. If $P\supseteq (S_1)$ we get $P\in V(S_1)$
and we are done. Hence, assume  $P\nsupseteq (S_1)$. Then there is a
 $y\in (S_1)$
such that $y\not\in P$. But now $y\cdot (S_2)$ is a subset of both $(S_1)$ and
$(S_2)$ because they are ideals. Hence, $y\cdot (S_2)\subseteq P$.
By the prime ideal condition $(S_2)\subseteq P$ which we had to show.\qed
\bigskip
\remark{Remark 1}
The closed points in $\Spec(R)$ are  the prime ideals which
are maximal ideals.
\endremark
\remark{Remark 2}
If we take any prime ideal $P$ then the (topological) closure of
$P$ in $\Spec(R)$ is given as
$$V(P)=\{\,Q\in \Spec(R)\mid Q\supseteq P\,\}\ .$$
Hence, the closure of $P$ consists of $P$ and all ``subvarieties'' of $P$
together. In particular the closure of a curve consists of the
curve as geometric object and  all points lying on the curve.
\endremark
\bigskip
At the end of this lecture let me return to the affine line over a
field $\K$, resp\. its algebraic model the polynomial ring in
one variable $\K[X]$.
We saw already that we have the non-closed point corresponding
to the prime ideal $\{0\}$ and the closed points corresponding to the
prime ideals $(f)$ (which are automatically maximal)
 where $f$ is an irreducible polynomial of
degree $\ge 1$. If $\K$ is an algebraically closed field the only
irreducible polynomials are the linear polynomials
$X-\a$. Hence, the closed points of $\ \Spec(\K[X])\ $ indeed correspond
to the geometric points  $\a\in \K$. The non-closed point corresponds to
the whole affine line.

Now we want to drop the condition that $\K$ is algebraically closed.
As example let us consider $\R[X]$.
We have two different types of irreducible polynomials.
Of type (i) are  the linear polynomials $X-\a$ (with a real zero $\a$) and
of type (ii) are the quadratic polynomials $\ X^2+2aX+b\ $ with pairs of
conjugate complex zeros.
The maximal ideals generated by the
polynomials of type (i) correspond again to the geometric points of
$\R$. There is no such  relation for  type (ii).
In this case we have $\ V(X^2+2aX+b)=\emptyset$. Hence, there is no
subvariety at all associated to this ideal. But if we calculate
the coordinate ring $R(A)$ of this (not existing) subvariety $A$ we
obtain
$$R(A)=\R[X]/(X^2+2aX+b)\cong \R\oplus \R\bar X$$
 with
the relation $\bar X^2=-2a\bar X-b$. In particular,
$R(A)$ is a two-dimensional vector space.
It is easy to show that $R(A)$ is isomorphic to $\C$.
Instead of describing the ``point'' $A$ as non-existing we should better
describe it as a point of the real affine line which is $\C-$valued.
(Recall that for the points of type (i) $R(A)\cong \R$.)
This corresponds to the fact that the polynomial splits
over the  complex numbers $\C$ into two factors
$$(X+(a+\sqrt{a^2-b}))(X+(a-\sqrt{a^2-b})).$$
In this sense, the ideals of type (ii) correspond to
conjugate  pairs of
complex numbers. Note that there is no way to distinguish between the
two numbers from our point of view.

In the general situation for $\K$ one has to consider $\Bbb L-$valued
points, where $\Bbb L$ is
allowed to be any finite-dimensional field extension
of $\K$.
\vskip 1cm
\head
\bf 3. Homomorphisms
\endhead
\vskip 0.7cm
\subheading{
Part 1}
Let $V$ and $W$ be algebraic sets (not necessarily irreducible),
resp\.
$$R(V)=\Kxn/I,\qquad R(W)=\Kym/J$$
their coordinate rings.
If $\Phi:V\to W$ is an arbitrary map and $f:W\to\K$ is a function then
the pull-back $\ \Phi^*(f):=f\circ \Phi\ $ is a function $\ V\ \to\K$.
If we interpret the elements of $R(W)$ as functions
we
want to call
$\Phi$ an {\sl algebraic map} if $\Phi^*(f)\in R(V)$ for every $f\in R(W)$.
Roughly speaking this is equivalent to the fact that
$\Phi$ ``comes'' from an algebra homomorphism
$R(W)\to R(V)$.
In this sense the coordinate rings are the dual objects to the
algebraic varieties.

To make this precise, especially  also  to take care
of the multiplicities, we should start from the other direction.
Let $\Psi:R(W)\to R(V)$ be an algebra homomorphism. This homomorphism
defines  a homomorphism $\widetilde{\Psi}$ (where $\nu$ is
the natural quotient map)
$$\widetilde{\Psi}=\Psi\circ\nu:\Kym\to R(V)\quad
\text{with}\quad\widetilde{\Psi}(J)=0 \mod I\ .$$
Such a homomorphism is given if we know the elements
$\widetilde{\Psi}(Y_j)$. Conversely, if we fix elements
$r_1,r_2,\ldots,r_m\in R(V)$ then
$\widetilde{\Psi}(Y_j):=r_j$, for $j=1,\ldots, m$
 defines an algebra homomorphism
$\ \widetilde{\Psi}:\Kym\to R(V)$.
If $f(r_1,r_2,\ldots,r_m)=0 \mod I$ for all $f\in J$ then
 $\widetilde{\Psi}$ factorizes
through $R(W)$.
Such a map indeed defines a map $\Psi^*$ on the set of geometric points,
$$\Psi^*:\ V\ \to\  W,\qquad \Psi^*(\a_1,\a_2,\ldots.\a_n)
:=(\b_1,\b_2,\ldots,\b_m)$$
where the $\b_j$ are defined as
$$\b_j=Y_j(\Psi^*(\a_1,\a_2,\ldots,\a_n)):=\widetilde{\Psi}(Y_i)
(\a_1,\a_2,\ldots,\a_n).$$
We  have  to check whether $\Psi^*(\a)=\b\in\K^m$ lies on the algebraic set
$W$ for $\a\in V$. For this we have to show that for all $f\in J$ we get
$\ f(\Psi^*(\a))=0$ for $\a\in V$ .
But
$$
f(\Psi^*(\a))=f\big(Y_1(\Psi^*(\a)),\ldots, Y_m(\Psi^*(\a))\big)=
f\big(\widetilde{\Psi}(Y_1)(\a),\ldots,\widetilde{ \Psi}(Y_m)(\a)\big)
=\widetilde{\Psi}(f)(\a)\ .
$$
Now $\ \widetilde{\Psi}(f)=0\ $, hence the claim.
\remark{Example 1}
A function $V\to \K$ is  given on the dual objects as
a $\K-$algebra homomorphism
$$\Phi:\K[T]\to
R(V)=\Kxn/I.$$
Such a  $\Phi$ is uniquely given by choosing an
arbitrary element $a\in R(V)$ and defining $\Phi(T):=a$.
Here again you see the (now complete) interpretation
of the elements of $R(V)$ as functions on $V$.
\endremark
\remark{Example 2}
The geometric process of
choosing  a (closed) point $\a$ on $V$ can alternatively be described as
giving a map from the algebraic variety consisting just of
one point to the variety.
Changing to the dual objects such a map is given as a map
$\Phi_\a$ from $R(V)$ to the field $\K$
which is the coordinate  ring of a point.
In this sense points correspond to homomorphisms of the coordinate ring
to the base field $\K$.
Such a homomorphism has of course a kernel $\ \ker \Phi_\a\ $ which is a
maximal ideal. Again, it is the ideal defining the closed point
$\a$.
\endremark
We will study this relation later. But first we take a different
look on the situation.
\subheading{Part 2}
Let $R$ be a $\K$-algebra where $\K$ is a field.
The typical examples are the quotients of the polynomial ring
$\Kxn$. Let $M$ be a module over $R$, i.e\.
a linear structure over $R$. In particular, $M$ is a vector space over
$\K$.
Some standard examples of modules are obtained in the following manner.
Let $I$ be an ideal of $R$, $\nu:R\to R/I$  the quotient
map then $\ R/I\ $ is a module over $R$ by defining
$r\cdot \nu(m):=\nu(r\cdot m)$.
\definition{Definition}
Let $M$ be a module over $R$. The {\sl annulator ideal}
is defined to be
$$\Ann(M):=\{\,r\in R\mid r\cdot m=0,\ \forall m\in M\}\ .$$
\enddefinition
\noindent
That $\Ann(M)$ is an ideal is easy to check.
It is also obvious  that $M$ is  a module over
$\ R/\Ann(M)$.
By construction in the above example  the ideal $I$
is the annulator ideal of $R/I$. Hence, every ideal of $R$
is the annulator ideal
of a suitable $R-$module.
\definition{Definition}
A module $M$ is called a {\sl simple} module if $M\ne \{0\}$ and $M$ has only
the
trivial submodules $\{0\}$ and $M$.
\enddefinition
\proclaim{Claim}
$M$ is  a simple module if and only if  there is a maximal  ideal $P$
such that $\ M\cong R/P$.
\endproclaim
\demo{Proof}
Note that the submodules of $R/P$ correspond to the ideals lying between
$R$ and $P$. Hence, if $P$ is maximal then $R/P$ is simple.
Conversely, given a simple module $M$
 take $m\in M, m\ne 0$. Then $R\cdot m$ is a submodule of
$M$. Because $1\cdot m=m$ the module $R\cdot m\ne \{0\}$, hence it is the
whole module $M$.
The map $\varphi(r)=r\cdot m$ defines a surjective map $\varphi :R\to M$.
This map is an $R-$module map where $R$ is considered as a module
over itself. The kernel $P$ of such a map is an $R-$submodule. But
$R-$submodules of $R$ are nothing else than ideals of $R$.
In view of the next lecture where we drop the commutativity
let us note already that
submodules of a ring $R$ are more precisely the left ideals of $R$.
The kernel $P$ has to be maximal otherwise the image of a maximal ideal lying
between $P$ and $R$ would be a non-trivial submodule of $M$.
Hence, $M\cong R/P$.
\qed
\enddemo
{}From this point of view the maximal ideals of $\ R(V)\ $ correspond to
$R(V)-$module homomorphisms to simple $R(V)-$modules.
If  $R(V)$ is a algebra over the field $\K$, then
a simple module $M$ is of course a vector space over $\K$.
By the above, we saw that it is even a field extension of $\K$.
(Recall that $M\cong R/P$ with $P$ a maximal ideal).
Because $R(V)$ is  finitely generated as $\K-$algebra
it is a  finite dimensional vector space over $K$
(see \cite{\KUKA,\rm S.56})
hence, a finite (algebraic) field extension.
\proclaim{Observation}
The maximal ideals (the ``points'')
 of $R=\Kxn/I$  correspond to the
$\K-$algebra homomorphism from $R$ to arbitrary finite (algebraic) field
extensions $\Bbb L$  of the base field $\K$.
We call these homomorphisms $\Bbb L-$valued points.
\endproclaim
In particular, if the field $\K$ is algebraically closed there are
no nontrivial algebraic field extensions. Hence,
there are only  $\K-$valued points.
If we consider  reduced varieties
(i.e\. varieties whose coordinate rings are reduced rings)
we get a complete dictionary.
Let $V$ be a variety, $P=I(V)$ the associated prime ideal
generated as $P=(f_1,f_2,\ldots,f_r)$ with $f_i\in\Kxn$
suitable polynomials, and
$R(V)$ the coordinate ring $R_n/P$.
\nl The points can be given in 3 ways:

\roster
\item[1] As classical points.
$\a=(\a_1,\a_2,\ldots,\a_n)\in\K^n$ with
\nl
$f_1(\a)=f_2(\a)=\cdots=f_r(\a)=0$.
\item[2]
As maximal ideals
in $R(V)$. They in turn can be identified
with the maximal ideals in $\Kxn$ which contain the prime ideal $P$.
In an explicit manner these can be given as
$\Mxn$ with the condition $f_1(\a)=f_2(\a)=\cdots=f_r(\a)=0$.
\item[3]
As surjective algebra homomorphisms $\ \phi:R(V)\to \K$. They are
fixed by defining $\bar X_i\mapsto \phi(\bar X_i)=\a_i,  i=1,\ldots, n$
 in such
a way that
\nl $\phi(f_1)=\phi(f_2)=\cdots=\phi(f_r)=0$.
\endroster

The situation is different if we drop the assumption that
$\K$ is algebraically closed.
The typical changes can already be seen if we take the real numbers
$\R$ and the real affine line.
The associated coordinate ring is $\R[X]$. There are only two
finite extension fields of $\R$, either $\R$ itself or the complex
number field $\C$. If we consider $\R-$algebra   homomorphism
from $\ \R[X]\ $ to $\ \C\ $ then they are given by
prescribing $\ X\mapsto \a\in\C$. If $\a\in\R$ we are again in
 the same situation as
above (this gives us the type (i) maximal ideals).
If $\a\not\in\R$ then the kernel $I$ of the map is a  maximal
ideal of type (ii) $I=(f)$ where $f$ is a quadratic polynomial.
$f$ has $\a$ and $\bar\a$ as zeros.
This says that the homomorphism $\Psi_{\bar \a}:X\mapsto\bar\alpha$
 which is clearly different
from $\Psi_{\a}:X\mapsto\alpha$ has the same kernel.
In particular,
for one maximal ideal of type (ii) we have two different
homomorphisms.
Note that the map $\a\to\bar\a$ is an element of the Galois group
$G(\C/\R)=\{id,\tau\}$ where $\tau$ is complex conjugation.
The two homomorphisms $\Psi_{\a}$ and $\Psi_{\bar\a}$
are related as $\  \Psi_{\bar\a}=\tau\circ \Psi_{\a}$.

This is indeed the general situation for $R(V)$, a finitely generated
$\K-$algebra.
In general, there is no 1-1 correspondence between (1) and (2) anymore.
But there is a 1-1 correspondence between maximal ideals of
$R(V)$ and orbits of $\K-$algebra homomorphism of $R(V)$
 onto finite field extensions $\Bbb L$ of $\K$ under the action of the
Galois group
$$G(\Bbb L/\K):=\{\,\sigma:\Bbb L\to \Bbb L\ \text{ an automorphism of fields
with\ }
\sigma_{|\K}=id\}\ .$$
\vskip 1cm
\head
\bf 4. Some Comments on the noncommutative situation
\endhead
\vskip 0.7cm
For the following let $R$ be a (not necessarily commutative) algebra
over the field $\K$.
First, we have to distinguish in this more general context
left ideals (e.g\. subrings $I$ which are invariant under
multiplication with $R$ from the left), right ideals and
two-sided ideals (which are left  and right ideals).
To construct quotient rings two-sided ideals are needed.
If we use the term ideal without any additional comment we assume
the ideal to be a two-sided one.

We want to introduce the concepts of prime ideals, maximal ideals, etc\..
A first definition of a prime ideal could be as follows.
We call a two-sided ideal I $prime$ if the quotient $R/I$
contains no zero-divisor. This definition has the drawback that
there are rings without any prime ideal at all.
Take for example the ring of $2\times 2$ matrices.
Beside the ideal $\{0\}$ and the whole ring the matrix ring does not
contain any other ideal.
To see this assume there is an ideal $I$ which
contains a non-zero matrix $A$. By applying
elementary operations from the left and the right we can
transform any
matrix to normal form which is
a diagonal matrix with just $1$ (at least one) and $0$ on the diagonal.
By multiplication with a
permutation matrix we can achieve any pattern in the diagonal.
These operations keep us inside the ideal.
Adding suitable elements  we see that the unit matrix is in the ideal.
 Hence the ideal is
the whole ring.
But obviously, the matrix ring has zero divisors. Hence, $\{0\}$ is not
prime in this definition. We see that this
ring does not contain any prime ideal at all with respect to
the definition.
We choose
another name for  such ideals: they are called {\sl complete prime ideals}.
\definition{Definition}
A (two-sided) ideal $I$ is called a {\sl prime ideal}
if for any two ideals $J_1$ and $J_2$ with $\ J_1\cdot J_2\subseteq I\ $
it follows that
$\ J_1\subseteq I\ $ or $\ J_2\subseteq I\ $.
\enddefinition
\noindent
This definition is equivalent to the following one.
\definition{Definition}
A (two-sided) ideal $I$ is called a {\sl prime ideal}
if for any two elements $a,b\in R$  with $\ a\cdot R\cdot b\subseteq I\ $
it follows that $a\in I$ or $b\in I$.
\enddefinition
\demo{Proof}
(2.~D)$\implies$ (1.~D): Take $J_1\nsubseteq I $ and $J_2\nsubseteq I$ ideals.
We have to show that $J_1\cdot J_2\nsubseteq I$.
For this choose $x\in J_1\setminus I$ and $y\in J_2\setminus I$.
Then $x\cdot R\cdot y\subseteq J_1\cdot J_2$ but there must be some $r\in R$
such that $x\cdot r\cdot y\not\in I$ due to the condition that
$I$ is prime with respect to (2.~D). Hence, $J_1\cdot J_2\nsubseteq I $
which is the claim.
\nl
(1.~D)$\implies$ (2.~D):
Take $a,b\in R$. The ideals generated by these elements are $RaR$
 and $RbR$. The
product of these "principal" ideals is not a principal ideal anymore.
It is $RaR\cdot RbR=RaRbR:=(arb\mid r\in R)$.
Assume $arb\in I$ for all $r\in R$. Hence
$\ (RaR)(RbR)\subseteq I\ $ and because $I$ is
prime we obtain by the first  definition (1.~D) that
either $RaR$ or $RbR$ are in $I$. Taking as  element of $R$ the $1$
we get $a\in R$ or $b\in R$.
\qed\enddemo
Every ideal which is a complete prime is prime.
Obviously, the condition (2.~D) is a weaker condition than the
condition that already from $a\cdot b\in I$
it follows that $a\in R$ or $b\in R$ (which is equivalent to:
$R/I$ contains no zero-divisors).
If $R$ is commutative then they coincide.
In this case $a\cdot r\cdot b=r\cdot a\cdot b$,
and with $a\cdot b\in I$ also $r\cdot a\cdot b\in I$ which is no
additional condition.
Here you see clearly where the noncommutativity enters the picture.
In the ring of matrices the ideal $\{0\}$ is prime because
if after fixing two matrices $A$ and $B$ we obtain $A\cdot T\cdot B=0$
 for any matrix $T$
then either $A$ or $B$ has to be the
zero matrix.
This  shows that the zero ideal in the matrix ring is a prime ideal.

{\sl Maximal ideals} are defined again as in the commutative setting
just as maximal elements in  the (non-empty) set of ideals.
By Zorn's lemma there exist maximal ideals.
\proclaim{Claim}
If $M$ is a maximal ideal then it is a prime ideal.
\endproclaim
\demo{Proof}
Take $I$ and $J$ ideals of $R$ which are not contained in $M$.
Then by the maximality of $M$ we get $(I+M)=R$ and $(J+M)=R$ hence,
$$R\cdot R=R=(I+M)(J+M)=I\cdot J+M\cdot J+I\cdot M+
M\cdot M\ .$$
If we assume $I\cdot J\subseteq M$ then $R\subseteq M$ which is
a contradiction. Hence $I\cdot J\nsubseteq M$. This shows $M$ is prime.
\qed
\enddemo
By this result we see that every ring has prime ideals.
\nl
In the commutative case if
we approach the theory of  ideals from the point of view
of  modules over $R$ we obtain an equivalent description. This is
not true anymore in the noncommutative setting.
For this let $M$ be a (left-)module over $R$. As above we define
$$\Ann(M):=\{\,r\in R\mid r\cdot m=0,\forall m\in M\}\ ,$$
the annulator of the module $M$.
$\Ann(M)$ is a two-sided ideal. Clearly, it is closed under addition
and is a left ideal. (This is even true for an annulator of a single element
$m\in M$). It is also a right ideal: let $s\in \Ann(M)$
 and $t\in R$ then $(st)m=s(tm)=0$
because $s$ annulates also $tm$.
\definition{Definition}
An ideal $I$ is called a {\sl primitive ideal} if
$I$ is the annulator ideal of a simple module $M$.
\enddefinition
Let us call the set of prime, resp\.  primitive, resp\. maximal
ideals $\ \Spec(R),\ \Priv(R)\ $ and $\ \Max(R)$.
\proclaim{Claim}
$$\Spec(R)\quad \supseteq\quad  \Priv(R)\quad \supseteq\quad  \Max(R)\ .$$
\endproclaim
\demo{Proof}
(1). Let $P$ be a maximal ideal. Then $R/P$ is a (left-)module.
Unfortunately, it is not necessarily simple (as module). The submodules
correspond to left-ideals lying between $P$ and $R$.
Choose $Q$ a maximal left ideal lying above $P$.
Then $R/Q$ is a simple (left-) module and $P\cdot (R/Q)=0$ because
$P\cdot R=P\subseteq Q$. Hence, $P\subseteq \Ann(R/Q)$ and because
$\Ann(R/Q)$ is a two-sided ideal we get equality.
\nl
(2). Take $P=\Ann(M)$, a primitive ideal. Assume $P$ is not prime.
 Then there exist
$a,b\in R$ but $a,b\not\in P$ such that for all $r\in R$ we get
$arb\in P$. This implies $arbm=0$ for all $m\in M$
but $bm\ne 0$ for at least one $m$.  Now $B=R(bm)$ is a non-vanishing
submodule.
Obviously, $a\in \Ann(B)$, hence $B\ne M$. This contradicts the
simplicity of $M$.
\qed\enddemo
Clearly, in the commutative case $\ \Priv(R)=\Max(R)$.
Let me just give an example from \cite{\GOWA} that in the noncommutative
case they fall apart.
Take $V$ an infinite-dimensional $\C-$vector space. Let $R$ be the
algebra  of linear endomorphisms of $V$ and $I$ the nontrivial
two-sided  ideal consisting of
linear endomorphisms with finite-dimensional image.
The vector space $V$ is an $R-$module by the natural action of the
endomorphisms. We get that
$V=R\cdot v$ where $v$ is any non-zero vector of $V$. This implies that
the module
$V$ is simple and that $\Ann(V)=\{0\}$. Hence $\{0\}$ is primitive,
but it is not maximal because $I$ is lying above it.
\bigskip
In the commutative case we
saw that we could interpret homomorphisms of the coordinate ring
(which is an algebra if we consider varieties over a base field)
into a field as points
of the associated space.
Indeed, it is possible to give such an interpretation also
 in the noncommutative setting.
Let me  give an example, for details see \cite{\MAQG}.
Let $M_q(2)$ for $q\in \C, q\ne 0$ be the (noncommutative)
 $\C-$algebra generated by $a,b,c,d$,
subject to the relations:
$$
\gathered
ab=\frac 1q\,ba,\qquad ac=\frac 1q\, ca,\qquad ad=da+\left(\frac
1q-q\right)bc,\\
\qquad bc=cb,\qquad bd=\frac 1q\, db,\qquad cd=\frac 1q \,dc\ .
\endgathered
\tag 4-1
$$
This algebra is constructed by first considering all possible words in
$a,b,c,d$. This defines the free noncommutative algebra of
this alphabet. Multiplication is defined by concatenation of the
words. Take the ideal generated by the expressions
\ (left-side) -- (right-side)\  of all
the relations (4-1) and build the quotient algebra.
Note that for $q=1$ we obtain the commutative algebra of
polynomial functions on the space of all $2\times 2$ matrices over $\C$.
In this sense the algebra  $M_q(2)$  represents the ``quantum
matrices'' as a ``deformation of the usual matrices''.
To end up with the {\sl quantum group} $Gl_q(2)$ we would have to
add another element for the formal inverse of the
quantum determinant $D=ad-\dfrac 1q\, bc$.
\footnote{There are other objects which carry also the name
quantum groups.}

Now let $A$ be another algebra.
We call a $\C-$linear algebra homomorphism
\nl $\Psi\in \Hom(M_q(2),A)$
an $A-${\sl valued point} of $M_q(2)$. It is called a {\sl generic
point} if $\Psi$ is injective.
Saying that a linear map $\Psi$ is an algebra homomorphism is
 equivalent to saying that the elements
$\ \Psi(a),\Psi(b),\Psi(c),\Psi(d)\ $ fulfill the same relations (4-1)  as
the $a,b,c$ and $d$.
One might interpret $\Psi$ as a point of the ``quantum group''.
But be careful, it is only possible to ``multiply'' the two matrices if
the images of the two maps
$$\Psi_1 \sim B_1:=\pmatrix a_1&b_1\\c_1&d_1\endpmatrix,\qquad
\Psi_2 \sim B_2:=\pmatrix a_2&b_2\\c_2&d_2\endpmatrix\ ,$$
lie in a common algebra $A_3$,
i.e\. $a_1,b_1,c_1,d_1\in A_1\subseteq A_3$ and
$a_2,b_2,c_2,d_2\in A_2\subseteq A_3$.
Then  we can multiply the
two matrices $\ B_1\cdot B_2\ $ as prescribed by the
usual matrix product and obtain another matrix $B_3$ with
coefficients $a_3,b_3,c_3,d_3\in A_3$.
This matrix defines only then a homomorphism of $M_q(2)$, i.e\. an
$A_3-$valued point
if $\Psi_1(M_q(2))$ commutes with  $\Psi_2(M_q(2))$ as subalgebras
of $A_3$. In particular, the product of $\Psi$ with itself is not
an $A-$valued point of $M_q(2)$ anymore. One can show that it is an
$A-$valued point of
$M_{q^2}(2)$.

Because in the audience there a couple people who
had and still have  their
share in developing the fundamentals of quantum groups
(the Wess-Zumino approach) there is no need to
give a lot of references on the subject. Certainly, these people know
it much better than I do. For the reader let me just quote one
article by Julius Wess and Bruno Zumino \cite{\WEQG}
where one finds references  for further study in this
direction.
Let me only give the following three references of books, resp\. papers
of Manin which are more connected to the theme of these lectures:
``Quantum groups and noncommutative geometry'' \cite{\MAQG},
``Topics in  noncommutative geometry''  \cite{\MANG}, and
``Notes on quantum groups and the quantum de Rham complexes''  \cite{\MADR}.

For the general noncommutative situation I like to recommend
Goodearl and Warfield, ``An introduction to noncommutative
noetherian rings'' \cite{\GOWA} and Borho, Gabriel, Rentschler,
``Primideale in Einh\"ullenden aufl\"osbarer Liealgebren''
\cite{\BOGARE}.
These books are still completely on the algebraic side of the theory.
For the algebraic geometric side there is still not very much available.
Unfortunately, I am also not completely aware of the very recent
developments of the theory. The reader may use
the  two articles \cite{\ART} and \cite{\ROS} as starting points
for his own exploration of the subject.
\head
\vskip 1cm
\bf 5. Affine schemes
\endhead
\vskip 0.7cm
Returning to the commutative setting
let $R$ be again a commutative ring with unit $1$.
We do not assume $R$ to be an
algebra over a field $\K$.
If we consider the theory of
differentiable manifolds the model manifold is $\R^n$.
Locally any arbitrary
manifold looks like the model manifold.
Affine schemes are the ``model spaces'' of algebraic geometry. General
schemes will locally look like affine schemes.
Contrary to the differentiable setting, there is not just one
model space but a lot of them.
Affine schemes are very useful generalizations of  affine varieties.
Starting from  affine varieties $V$ over a field $\K$ we saw that we
were able to assign  dual objects to them, the coordinate rings $R(V)$.
The geometric structure of $V$ (subvarieties, points, maps, ...)
are represented by the algebraic structure of $R(V)$ (prime ideals, maximal
ideals, ring homomorphisms, ...).
After dualization we are even able to extend our notion of ``space''
in the sense that we can consider more general rings and regard
them as dual objects of some generalized ``spaces''.
In noncommutative quantum geometry one even studies certain
noncommutative algebras over a field $\K$. Quantum spaces are
the dual objects of these algebras.
We will restrict ourselves to the commutative case, but we will allow
arbitrary rings.

What are the dual objects (dual to the rings) which generalize the concept
of a variety. We saw already that prime ideals of the coordinate ring
correspond to subvarieties and that closed prime ideals (at least
if the field $\K$ is algebraically closed)  correspond to points.
It is quite natural to take as space the set $\ \Spec(R)\ $ together
with its Zariski topology. But this is not enough.
If we take for example $R_1=\K$ and $R_2=\K[\epsilon]/(\epsilon^2)$
then in both cases $\Spec(R_i)$ consists just of one point.
It is represented in the first case by the ideal $\{0\}$ in the second case
by $(\epsilon)$. Obviously, both $\ \Spec\ $ coincide.
Let us compare this with the differentiable setting.
For an arbitrary differentiable manifold the structure is not
yet given if we consider the manifold just as a topological manifold.
We can  fix its differentiable structure if we tell what
the differentiable functions are.
The same is necessary in the algebraic situation.
Hence, $\Spec(R)$ together with the functions (which
in the case of varieties correspond to the elements of $R$) should
be considered as ``space''.
So the space associated to a ring $R$ should be $\ (\Spec(R),R)$.
In fact, $\Spec(R)$ is not a data independent of $R$.
Nevertheless, we will write both information
in view of  globalizations of the notion.
 Compare this
again with the differentiable situation. If you have a manifold which is
$\R^n$ (the model manifold) then the topology is fixed.
But if you have an arbitrary differentiable manifold then you
need  a topology at the first place to define coordinate charts at all.
In view of these globalizations we
additionally  have to replace the ring of functions
by a data which will give us all local and global functions together.
Note that in the case of compact complex analytic manifolds there  would
exist no non-constant analytic functions at all.
The right setting for this is the language of sheaves.
Here it  is not the time and place to introduce this language.
Just let me give you a very rough idea. A sheaf is the coding of an
object which is local and global in a compatible way.
A standard example
(which is in
 some sense too simple) is the sheaf of differentiable functions on
a differentiable manifold $X$. It assigns to every open set
$U$ the ring of differentiable functions defined on $U$.
The compatibility just means that this assignment is compatible with
the restriction of the sets where the functions are defined on.
In Appendix~A to this lecture you will find the exact definition
of a sheaf of rings.
So, given a ring $R$ its associated affine scheme is the pair
$\ (\Spec(R),\Or)\ $ where $\ \Spec(R)\ $ is the
set of prime ideals made into a topological space
by  the Zariski topology and $\Or$ is a sheaf of
rings on $\Spec(R)$ which we will define in a minute.
For simplicity  this pair is sometimes just called $\Spec(R)$.

Recall that the sets $V(S):=\{\,P\in \Spec(R)\mid P\supseteq S\}$,
where $S$ is  any $S\subseteq R$,
are the closed  sets. Hence the  sets $\ \Spec(R)\setminus V(S)\ $ are
exactly the open sets of $X:=\Spec(R)$.
There are some special open sets in $X$. For a single element
$f\in S$ we define
$$X_f:=\Spec(R)\setminus V(f)=\{\,P\in \Spec(R)\mid f\not\in P\}\ .\tag 5-1$$
The set $\ \{X_f,f\in S\}\ $ is a basis of the topology
which  says that every open set is a union of  $X_f$.
This is especially useful because the $X_f$ are again  affine schemes.
More precisely,
$X_f=\Spec(R_f)$.
Here the ring $R_f$ is defined as the ring of fractions
with the  powers of $f$ as denominators:
$$R_f:=\{\,\frac g{f^n}\mid g\in R,\ n\in\N_0\;\}\ .$$
Let me explain this construction. It is a generalization of
the way how one constructs the rational numbers from
the integers. For this let $S$ be a multiplicative system, i.e\. a
subset of $R$ which is multiplicatively closed and contains 1.
(In our example, $S:=\{1,f,f^2,f^3,\ldots\}$.)
Now introduce on the set of  pairs in $\ R\times S\ $ the
equivalence relation
$$(t,s)\sim (t',s')\iff \exists s''\in S\text{ such that }
s''(s't-st')=0\ .$$
The equivalence class of $(s,t)$ is denoted by $\dfrac st$.
There is always a map $R\to R_f$ given by $r\mapsto \dfrac r1$.
The  ideals in $R_f$ are obtained by
mapping   the ideals $I$ of $R$ to $R_f$ and multiplying them
by $R_f$ : $R_f\cdot I$. By construction, $f$ is a unit in $R_f$.
Hence, if $f\in P$ where  $P$ is a prime ideal
then $R_f=R_f\cdot P$. If $f\not\in P$ then
$R_f\cdot P$ still is a prime ideal of $R_f$.
This shows  $\ X_f=\Spec(R_f)$. For details see \cite{\KUKA}.

You might ask what happens if $f$ is nilpotent, i.e\. if there is a
$n\in\N$ such that $f^n=0$. In this case $f$ is contained in any prime ideal
of $R_f$. Hence $\Spec(R_f)=\emptyset$  in agreement with
$R_f=\{0\}$.

If $f$ is not a zero divisor  the map $R\to R_f$ is an embedding
and if $f$ is not a unit in $R$ the ring $R_f$ will be bigger.
This is completely in accordance with our understanding of
$R$ resp\. $R_f$ as functions on $X$, resp\. on the honest
subset $X_f$.
Passing from $X$ to $X_f$ is something like passing from the global to
the more local situation.
This explains why this process of taking the ring
of fractions with respect to some multiplicative subset $S$
is sometimes called localization of the ring.
The reader is adviced to consider the following example.
Let $P$ be a prime ideal, show that $S=R\setminus P$ is a multiplicative
set. How can one interpret the ring of fractions of $R$ with
respect to $S$?

Now we define our sheaf $\Or$ for the     basis sets $X_f$.
In $\ X_f\cap X_g\ $ are the prime ideals which neither contain
$f$ nor $g$. Hence they do  not contain $f\cdot g$.
It follows that
$X_f\cap X_g=X_{fg}$.
We see that the set of the $X_f$ are closed under intersections.
Note also that $X_1=X$ and $X_0=\emptyset$. We define
$$\Or(X):=R,\qquad \Or(X_f):=R_f\ .\tag 5-2$$
For $X_{fg}=X_f\cap X_g\subseteq X_f$ we define the restriction map
$$\rho_{fg}^f:R_f\to (R_f)_g=R_{fg},\quad r\mapsto \frac r1\ .$$
It is easy to check
 that all the maps $\rho_{..}^{..}$ are compatible
on the intersections of the basis open sets.
In Appendix ~B I will show that the other sheaf axioms are  fulfilled  for the
$X_f$ with respect to their intersections.
Hence,  we have defined the sheaf $\Or$ on  a basis of
the topology which is closed under intersections.
The whole sheaf is now defined by some general construction.
We set
$$\Or(U):=\projlim_{X_f\subseteq U}\Or(X_f)$$
for a general open set. For more details  see \cite{\EHS}.
Let us collect the facts.
\definition{Definition}
Let $R$ be a commutative ring. The pair $\ (\Spec(R),\Or)$,
where $\Spec(R)$ is the space of prime ideals with the Zariski topology
and $\Or$ is the sheaf of rings on $\Spec(R)$ introduced above is
called the {\sl associated affine scheme} $\Spec(R)$ of $R$.
The sheaf $\Or$ is called the {\sl structure sheaf} of $\Spec(R)$.
\enddefinition
\bigskip
Let me explain in which sense the elements $f$ of
an arbitrary ring  $R$  can be considered as functions, i.e\. as
prescriptions how to assign a value from a field to every point.
This gives me the opportunity to introduce another important
concept which is related to points: the residue  fields.
Fix an element $f\in R$.
Let $[P]\in \Spec(R)$ be a (not necessarily closed) point,
 i.e\. $P$ is a prime ideal.
We define
$$f([P]):=f \mod P\in R/P\ $$
in a first step. From the primeness of $P$
it follows that $R/P$ is an integral domain  ring (i.e\. it
contains no zero-divisor). Hence $S:=(R/P)\setminus \{0\}$ is a
multiplicative system and the ring of fractions, denoted
by $\Quot(R/P)$, is a field, the quotient field.
Because $R/P$ is an integral domain  it can be embedded
into its quotient field. Hence, $f([P])$ is indeed an element of a field.
Contrary to the classical situation, if we change
the point $[P]$ the field $\Quot(R/P)$ will change too.

\remark{Example 1}
Take again $R=\Cxy$ and $f\in R$. Here we have three
 different types of points
in $\Spec(R)$.
\nl Type (i): the closed points $[M]$ with
$M=(X-\a,Y-\b)$ a maximal ideal.
We write
$\ f= f_0+(X-\a)\cdot g+(Y-\b)\cdot h\ $ with $f_0=f(\a,\b)\in\C$
and  $g,h\in R$.
Now
$$f([M])=f\mod M=f_0+(X-\a)\cdot g+(Y-\b)\cdot h\mod M=f_0\ .$$
The quotient $R/M$ is already a field, hence it is
the residue  field. In our case it is even the
base field $\C$.
The value $f([M])$ is just the value we obtain by plugging
the point $(\a,\b)$ into the polynomial $f$.
Note that the points are subvarieties of dimension 0.
\nl
Type (ii): the points $[P]$ with $P=(h)$, a principal ideal.
Here $h$ is an irreducible polynomial in the variables $X$ and $Y$.
If we calculate $R/P$ we obtain $\C[X,Y]/(h)$ which is not a field.
As residue  field we obtain $\C(X,Y)/(h)$. This field consists
of all rational expressions in the variables $X$ and $Y$ with the
relation $h(X,Y)=0$. This implies that the transcendence degree of
the residue  field over the base field is one, i.e\.
one of the variables $X$ or $Y$  is algebraically independent over $\C$ and
the second variable  is in an algebraic relation with the first and
the elements of $\C$. Note that the
coordinate ring has (Krull-) dimension one and
the subvariety corresponding to $[P]$ is a curve, i.e\. is an
object of  geometric dimension one.
\nl
Type (iii): [\{0\}] the zero ideal. In this case $R/P=\Cxy$ and
the residue field is $\C(X,Y)$ the rational function field
in two variables. In particular, its transcendence degree is two and
coincides with the (Krull-)dimensions of the coordinate ring and
the geometric dimension of the variety $V(\{0\})$ which equals
the whole affine plane $\C^2$.

Strictly speaking, we have not shown (and will not do it here) that
there are no other prime ideals. But this is in fact true,
see \cite{\KUKA}.
The equality of the transcendence degree
of the residue field and the
(Krull-) dimension of the coordinate ring obtained above
is true for all varieties over arbitrary fields.
For example, if we  replace $\C$ by $\R$ we  obtain for the
closed points, the maximal ideals, either $\R$ or $\C$ as residue
 fields. Both fields have transcendence degree 0 over $\R$.
\endremark
\remark{Example 2}
Consider $R=\Z$, the integers, then $\Spec(\Z)$ consists
of the zero ideal and the  principal ideals generated by
prime numbers.
As residue  field we obtain for $[{0}]$ the field
$\ \Quot(\Z/(0))=\Q\ $ and for the  point $[(p)]$  (which is a closed point)
$\F_p=\Z/(p)$, the  prime field of characteristic  $p$.
In particular, we see at this example that even for the maximal
points the residue  field can vary in an essential way.
Note that $\Z$ is not an  algebra over a fixed base field.
\endremark
\bigskip
Up to now we  considered one ring, resp\. one scheme.
In any category of objects one has  maps between the objects.
Let $\ \Phi:R\to S\ $ be a ring homomorphism. If $I$ is any ideal of
$S$, then $\Phi^{-1}(I)$ is an ideal of $R$.
The reader is  advised to check that if $P$ is prime
then $\Phi^{-1}(P)$ is again prime.
Hence, $\Phi^*:P\mapsto \Phi^{-1}(P)$ is a well-defined map
$\ \Spec(S)\to \Spec(R)$.
Indeed, it is even continuous because the pre-image of
a closed set is again closed.
Let $\ X=(\Spec(S),\Os)\ $ and $\ Y=(\Spec(R),\Or)\ $
be two affine schemes.
The map $\Phi$ induces also a map on the level of the structure sheaves
$\Phi_*:\Or\to\Os$.
The pair $(\Phi^*,\Phi_*)$ of maps fulfills certain compatibility
conditions which makes them to a homomorphism of schemes.

We will not work with schemes in general later on but let me give
at least for completeness the definition here.
\definition{Definition}
(a) A {\sl scheme} is a pair $\ X=(|X|,\Ox)\ $ consisting of a topological
space $|X|$ and a sheaf $\Ox$ of rings on $X$, such that
$X$ is locally  isomorphic to affine schemes
$(\Spec(R),\Or)$. This says that
for every point $x\in X$ there is an open set $U$ containing $x$,
and a ring $R$ (it may depend on the point $x$)
such that  the affine scheme $(\Spec(R),\Or)$ is
isomorphic to the scheme $(U,{\Cal O_{X|U}})$.
In other words there is a homeomorphism
$\Psi:U\to \Spec(R)$ such that there is an isomorphism of sheaves
$$\Psi^{\#}:\Psi_*(\Ox_{|U})\cong \Or\ .$$
Here the sheaf $\Psi_*(\Ox_{|U})$ is defined to be the sheaf on $\Spec(R)$
 given by the
assignment
$$
\Psi_*(\Ox_{|U})(W):=\Ox(\Psi^{-1}(W)),\qquad
\text{for every open set } W\subseteq \Spec(R) .$$
(b) A scheme is called an {\sl affine scheme} if it is globally isomorphic to
an affine scheme $(\Spec(R),\Or)$ associated to a ring $R$.
\enddefinition
\proclaim{Fact}
The category of affine schemes is equivalent to the category
of commutative rings with unit with the arrows
(representing the maps)  reversed.
\endproclaim

\bigskip
There are other important concepts in this theory.
First, there is the concept of a scheme over another scheme.
This is the right context to describe families of schemes.
Only within this framework it is possible to make such
useful things precise as
degenerations, moduli spaces etc.
Note that every affine scheme is in a natural
 way a scheme over $\ \Spec(\Z)$,
because for every ring $R$ we have the natural map
$\Z\to R,\ n\mapsto n\cdot 1\ $.
Taking the dual map introduced above we obtain a homomorphism of schemes.
\nl
If $R$ is a $\K-$algebra with $\K$ a field then we have the map
$\K\to R,\ \alpha\mapsto \alpha\cdot 1$, which is a ring homomorphism. Hence,
 we
 always obtain  a map:
$\Spec(R)\to \Spec(\K)=(\{0\},K)$.
By considering the coordinate ring $R(V)$ of an affine variety $V$ over
a fixed algebraically closed field $\K$ and assigning to it
the affine scheme $\Spec(R(V))$ we obtain a functor
from the category of varieties over $\K$ to the
category of schemes over $\K$. The schemes corresponding
 to the varieties are  the
{\sl irreducible   and reduced noetherian affine schemes of finite
type over $\Spec(\K)$}.
The additional properties of the scheme are nothing else as the
corresponding properties for the defining ring $R(V)$.
Here finite type means that $R(V)$ is a finitely generated $\K-$algebra.
You see again in which sense the schemes extend our
geometric objects from the varieties to more general ``spaces''.

The second concept is the concept of a {\sl functor of points }of a scheme.
We saw already at several places in the lectures that points of
a geometric object can be described as
homomorphisms of the dual (algebraic) object into some
simple (algebraic) object.
If $X$ is a scheme we can associate to it the following
functor from the category of schemes to the category of
sets: $h_X(S)=\Hom(S,X)$. Here $S$ is allowed to be any
 scheme and $\Hom(S,X)$ is the
set of homomorphisms of schemes
from $S$ to the fixed scheme $X$. Such a homomorphism is called an $S-${\sl
valued
point} of $X$. Note that we are in the geometric category, hence the order
of the elements in $\ \Hom(.,.)\ $ is just the other way round compared
to the former lectures.
The functor $h_X$ is called the {\sl functor of points} associated to
$X$.
Now $X$ is completely fixed by the functor $h_X$. In categorical
language: $X$ represents its own functor of points.
The advantage of this view-point is that certain
questions of algebraic geometry, like the existence
of a moduli space for certain geometric data, can be easily
transfered to the language of functors. One can extract already
a lot of geometric data without knowing whether there is
indeed a scheme having this functor as functor of points
(i.e\. representing the functor).
If you want to know more about this beautiful subject you should consult
\cite{\EHS} and \cite{\MPMP}.
\bigskip
\subhead
Appendix~A: The definition of a sheaf of rings
\endsubhead
A {\sl presheaf} $\Cal F$ of rings over a topological
space $X$ assigns to every open set $U$ in $X$
a ring $\Cal F(U)$ and to every pair of open sets $\ V\subseteq U\ $
a homomorphism of rings
$$\rho^U_V:\Cal F(U)\to\Cal F(V),$$
(the so called {\sl restriction map}) in such a way that
$$\gather \rho^U_U=id,\\
\rho^U_V\circ\rho^W_U=\rho^W_V\quad \text{ for }\ V\subseteq U\subseteq W\ .
\endgather$$
Instead of $\rho^U_V(f)$ for $f\in\Cal F(U)$ we often use the simpler
notation $f_{|V}$.
A presheaf is called a {\sl sheaf} if
for every open set
$U$ and every covering $(U_i)$ of this open set we have in addition:
\nl (1) if $ f,g\in\Cal F(U)$ with
$$f_{|U_i}=g_{|U_i}$$ for all $U_i$ then $f=g$,
\nl (2) if a set of $f_i\in\Cal F(U_i)$ is given
with $$f_{i|U_i\cap U_j}=f_{j|U_i\cap U_j} $$then there
exists  a $f\in\Cal F(U)$ with
$$f_{|U_i}=f_i.$$
Given two sheaves of rings $\Cal F$ and $\Cal G$ on $X$.
By a {\sl sheaf homomorphism}
$$\psi:\Cal F\to\Cal G$$
we understand an assignment of a ring homomorphism $\psi_U$
(for  every open set $U$)
$$\psi_U:\Cal F(U)\to\Cal G(U),$$
which is compatible with the restriction homomorphisms
$$\matrix U&&&\quad\Cal F(U)&\mapright{\psi_U}&\Cal G(U)\\
&&&&\\ \bigcup&&&\quad\mapdown{}&&\mapdown{}\\&&&&
\\V&&&\quad\Cal F(V)&\mapright{\psi_V}&\Cal G(V)\endmatrix
\qquad\qquad$$
More information you find in \cite{\SCHLRS}.
\subhead
Appendix B.  The structure sheaf $\Or$
\endsubhead
In this appendix I like to show that the sheaf axioms for the
structure sheaf $\Or$ on $X=\Spec(R)$ are fulfilled if we consider
only the basis open sets $\ X_f=\Spec(R)\setminus V(f)\ $.
Recall that the intersection of two basis basis open sets
$\ X_f\cap X_g=X_{fg}\ $ is again a basis open set.
The sheaf $\Or$ on the basis open sets was defined to be
$\ \Or(X_f)=R_f$ and the restriction maps
were the natural maps
$$R_f\ \to\ (R_f)_g\ = \ R_{fg},\qquad r\mapsto \frac r1\ .$$
Here I am following very closely the presentation in
\cite{\EHS}.
\proclaim{Lemma 1}
The set $\ \{X_f\mid f\in R\}\ $ is a basis of the topology.
\endproclaim
\demo{Proof}
We have to show that every open set $U$ is a union of such $X_f$.
By definition,
$$U=\Spec(R)\setminus V(S)=\Spec(R)\setminus(\bigcap\limits_{f\in S} V(f))
=\bigcup_{f\in S}(\Spec(R)\setminus V(f))=\bigcup_{f\in S}X_f\ .\qed$$
\enddemo
Obviously, only a set of generators $\ \{f_i\mid i\in J\}\ $ of
the ideal generated by the set $S$ is needed. Hence, if $R$ is a noetherian
ring every open set can already be covered by finitely manx $X_f$.
\proclaim{Lemma 2}
Let $\ X=\Spec(R)$ and $\ \{f_i\}_{i\in J}\ $ a set of elements of $R$
then the union of the sets $X_{f_i}$ equals $X$ if and only if the ideal
generated by the $f_i$ equals the whole ring $R$.
\endproclaim
\demo{Proof}
The union of the $X_{f_i}$ covers $\Spec(R)$ iff no prime ideal of
$R$ contains all the $f_i$. But every  ring strictly smaller
than the whole ring is dominated by a maximal (and hence prime)
ideal. The above can only be the case iff the ideal generated by the
$f_i$ is the whole ring.
\qed
\enddemo
\proclaim{Lemma 3}
The affine scheme  $\ X=\Spec(R)$ is a quasicompact space.
This says every open cover of $X$ has a finite subcover.
\endproclaim
\demo{Proof}
Let $\ X=\bigcup\limits_{j\in J} X_j\ $ be a cover of $X$. Because the
basis open set $X_f$ are a basis of the topology, every $X_j$ can
 be given as union of $X_{f_i}$. Altogether, we get a refinement
of the cover  $\ X=\bigcup\limits_{i\in I} X_{f_i}\ $.
By Lemma 2 the ideal generated by
these $f_i$ is the whole ring. In particular, 1 is a finite
linear combination of the $f_i$. Taking only these $f_i$ which occur with
a non-zero coefficient in the linear combination we get
(using Lemma 2 again) that
$\ X_{f_{i_k}},\ k=1,..,r$ is a finite subcover of $X$. Taking for every
$k$ just one element $X_{j_k}$ containing $X_{f_{i_k}}$ we obtain a
finite number of sets which is a subcover from the cover we started with.
\qed
\enddemo
Note that this space is not called a compact space because
the Hausdorff condition that every distinct two points
have disjoint open neighbourhoods is obviously not fulfilled.

The following proposition says that
the sheaf axioms (1) and (2) from App.~A for the basis open sets are
fulfilled.
\proclaim{Proposition}
Let $X_f$ be coverd by $\ \{X_{f_i}\}_{i\in I}$.
\vskip 0.2cm
(a) Let $g,h\in R_f=\Or(X_f)$ with $\ g=h\ $ as elements in
$\ R_{f_i}=\Or(X_{f_i})\ $ for every $i\in I$, then
$\ g=h\ $ also in $R_f$.
\vskip 0.2cm
(b)
Let $g_i\in R_{f_i}\ $ be given for all $i\in I$ with
$\ g_i=g_j\ $ in $R_{f_if_j}\ $, then there exist a $\ g\in R_f$ with
$\ g= g_i\ $ in $R_{f_i}$.
\endproclaim
\demo{Proof}
Because $X_f=\Spec (R_f)$ is again an affine scheme it is enough to
show the proposition for $R_f=R$, where $R$ is an arbitrary ring.
Let $\ X=\bigcup\limits_{i\in I}X_{f_i}\ $.
\vskip 0.2cm
(a) Let $ g,h\in R$ be such that they map to the same element in $R_{f_i}$.
This can only be the case if in $R$ we have
$$f_i^{n_i}\cdot(g-h)=0, \qquad \forall i\in I,$$
(see the construction of the ring of fractions above).
Due to the quasicompactness it is enough to consider
finitely many $f_i$, $i=1,..,r$. Hence, there is a $N$ such that
for every $i$ the element $f_i^N$ annulates $(g-h)$. There is
another number $M$, depending on $N$ and $r$, such that we have for the
following ideals
$$
(f_1^N,f_2^N,\ldots,f_r^N)\supseteq
{(f_1,f_2,\ldots,f_r)}^M\ .$$
Because the $X_{f_i}$, $i=1,..,r$ are a cover of $X$ the ideal on
the right side equals $(1)$. Hence, also the ideal on the left.
Combining 1 as linear combination of the generator we get
$$1\cdot(g-h)=(c_1f_1^N+c_2f_2^N\cdots+c_rf_r^N)(g-h)=0\ .$$
This shows (a)
\nl
(b) Let $g_i\in R_{f_i}$, $i\in I$ be given such that
$g_i=g_j$ in $R_{f_if_j}$. This says there as a $N$ such that
$${(f_if_j)}^Ng_i={(f_if_j)}^Ng_j$$
in $R$. Note that every $g_i$ can be written as $\dfrac {g_i^*}{f_i^{k_i}}$
with $g_i^*\in R$. Hence, if $N$ is big enough the elements
$f_i^Ng_i$  are in $R$.
Again by the quasicompactness a common $N$ will do it for every pair
$(i,j)$. Using the same arguments as in (a) we get
$$1\ = \ \sum e_if_i^N,\qquad e_i\in R\ .$$
This formula corresponds to a ``partition of unity''.
We set
$$g\ =\ \sum e_if_i^Ng_i\ .$$
We get
$$f_j^Ng=\sum_if_j^Ne_if_i^Ng_i=\sum_ie_if_i^Nf_j^Ng_j=
f_j^Ng_j\ .$$
This shows $\ g=g_j$ in $R_{f_j}$.
\qed
\enddemo
\np
\vskip 1cm
\head
\bf 6. Examples of Schemes
\endhead
\vskip 0.7cm
\subhead
1.~Projective Varieties
\endsubhead
Affine Varieties are examples of affine schemes
over a field $\K$. They have been covered thoroughly in
the other lectures. For completeness let me mention that it is
 possible to introduce the {\sl projective space}
$\ \P^n_{\K}\ $ of dimension
$n$ over a field $\K$. It can be given as orbit space
$(\K^{n+1}\setminus\{0\})/ \sim$, where two $(n+1)-$tuple $\a$ and $\b$ are
equivalent if $\a=\l\cdot \b$ with $\l\in \K$, $\l\ne 0$.
{\sl Projective varieties} are defined to be  the vanishing sets of
homogeneous polynomials in $n+1$ variables. See for example
\cite{\SCHLRS} for more information.
What makes them so interesting is that they are compact
varieties (if $\K=\C$ or $\R$).
Again everything can be dualized. One considers the projective coordinate
ring and its set of homogeneous ideals (ideals which are generated by
homogeneous elements). In the case of $\ \P^n_\K\ $ the homogeneous
coordinate ring is $\K[Y_0,Y_1,\ldots,Y_n]$.
Again it is possible to introduce
the Zariski topology on the set of  homogeneous prime ideals.
It is even possible to introduce the notion of a projective scheme
$\Proj$, which is again a topological space together with
a sheaf of rings, see \cite{\EHS}.

In the same way as $\ \P^n_\K\ $ can be covered by $(n+1)$ affine spaces
$\K^n$ it is possible to cover every   projective scheme by finitely many
affine schemes. This covering is even such that the projective
scheme is locally isomorphic  to
these affine scheme. Hence, it is a scheme.
The projective scheme $\ \Proj(\K[Y_0,Y_1,\ldots,Y_n])\ $
is locally isomorphic to $\ \Spec(\Kxn)\ $. For example,
the open set of elements $\a$ with  $Y_0(\a)\ne 0$ is in 1-1 correspondence
to it via the assignment $X_i\mapsto\dfrac {Y_i}{Y_0}$.

As already said, the projective schemes are schemes and you might
ask why should one pay special attention to them.
Projective schemes are quite
useful. They are schemes with rather strong additional
properties. For example, in the classical case (e.g\. nonsingular varieties
over
$\C$) projective varieties are compact in the classical
complex topology. This yields all the interesting  results like,
there are no non-constant global analytic or harmonic functions,
the theorem of Riemann-Roch is valid, the integration is well-defined,
and so on. Indeed, similar results we get for projective schemes.
Here it is the  feature ``properness'' which generalizes
compactness.
\subhead
2.~The scheme of integers
\endsubhead
The affine scheme $\Spec(\Z)=(\Spec(\Z),{\Cal O}_{\Spec(\Z)})$ we discussed
already in the last lecture.
The topological space consist of the element $[\{0\}]$ and
the elements $[(p)]$ where $p$ takes every prime number.
The residue  fields are $\Q$, resp\. the finite fields $\ \F_p$.
What are the closed sets. By definition, these are exactly the
sets $V(S)$ such that
there is a $S\subseteq \Z$ with
$$ V(S)\ :=\ \{\,[(p)]\in\Spec(\Z)\mid (p)\supseteq S\}\ =\  V((S))
\quad=\quad V((gcd(S)))\ .$$
For the last identification recall that the ideal
$(S)$ has to be generated by one element $n$ because
$\Z$ is a principle ideal ring. Now every element in $S$ has to be a multiple
of this $n$. We have to take the biggest such $n$ which
fulfills this condition,
hence $n=gcd(S)$.
If $n=0$ then $V(n)=V(0)=\Spec(\Z)$,
if $n=1$ then $V(n)=V(1)=\emptyset$, otherwise $V(n)$ consists of the finitely
many primes, resp\. their ideals, dividing $n$.
Altogether we get that the closed sets are    beside the whole space
and the empty set just sets of finitely many points. As already said
at some other place of these
lectures $\Z$ resembles very much $\K[X]$.
By the way, we see  that the
topologial closure $\overline{[\{0\}]}=\Spec(\Z)$ is the whole
space. For this reason $[\{0\}]$ is called the generic point of $\Spec(\Z)$.

All these has important consequences. We have two principles which
can be very useful:

(1.) Let some property be defined over $\Z$ and assume
it is a closed property. Assume further that the property is true for
infinitely many primes (e.g\. the property is true
if we consider the problem in characteristic $p$ for infinitely
many $p$) then it has to be true for the whole $\Spec (\Z)$.
Especially, it has to be true for all primes and
for the generic point, i.e\. in characteristic zero.

(2.) Now assume that the property is   an open property. If it is
true for at least one point, then it is true for all points
except for possibly
finitely many points. In particular, it has to be true for the
generic point (characteristic zero) because every non-empty open set
has to contain the generic point.

\subhead
3. A family of curve
\endsubhead
This example illustrates the second principle above.
To allow you to make further studies
by yourself
on the example I take the example from \cite{\EHS}.
You are encouraged to develop your own examples.
Consider the conic
$X^2-Y^2=5$. It defines a curve in the real (or complex) plane. In fact,
it is already defined over the integers which says nothing more than that
there is  a defining equation for the curve with integer coefficients.
Hence, it makes perfect
sense to ask for points $(\a,\b)\in\Z^2$ which solve the equation.
We already saw that it is advantagous to consider the coordinate ring.
The coordinate ring and everything else make sense also if there
would be no integer solution at all.
Here we have:
$$\Z\ \to\  R=\Z[X,Y]/(X^2-Y^2-5),\qquad
\Spec(R)\ \to\  \Spec(\Z)\ .$$
We obtain an affine scheme over $\Z$. Now $\Spec(\Z)$ is a one-dimensional
base, the fibres are one-dimensional curves, and $\Spec(R)$ is two-dimensional.
It is an arithmetic surface.
We want to study the fibres in more detail.
 Let $Y\to X$ be  a homomorphism of schemes
and $p$ a point on the base scheme $X$. The topological fibre over $p$ is just
the
usual pre-image of the point $p$. But here  we have to give the
fibre the structure of a scheme.
The general construction is as follows.
Represent the point $p$ by its residue
field $\ k(p)\ $ and a homomorphism of schemes
$\ \Spec (k(p))\to X\ $.
Take the ``fibre product of schemes'' of the scheme $\ Y\ $ with
$\ \Spec (k(p))\ $ over $X$.
Instead of giving the general definition
let me just write this down in our affine situation:
$$
\CD
\Spec(R)@<<< \Spec(R\bigotimes\limits_{\Z}k(p))@.
\qquad\qquad @. R@>>>R\bigotimes\limits_{\Z}k(p)\\
@VVV @VVV @. @AAA @AAA\\
\Spec(\Z)@<<< \Spec(k(p))@.\qquad\qquad  @. \Z@>>>k(p)
\endCD
$$
Both diagrams are commutative diagrams and are dual to each other.

Here we obtain for the generic point $[0]$ the residue field $\ k(0)=\Q\ $
and as fibre the $\Spec$ of
$$
 R{\tsize\bigotimes\limits_{\Z}}\Q=
\Q[X,Y]/(X^2-Y^2-5)\ .
$$ For the closed points $[p]$ we get $\ k(p)=\F_p\ $ and as  fibre the $\Spec$
of
$$
R{\tsize\bigotimes\limits_{\Z}}\F_p=
\F_p [X,Y]/(X^2-Y^2-5)\ .
$$
In the fibres  over the primes we just do calculation modulo $p$.
A point
lying on a curve in the plane is a singular point of the curve
if both
partial derivatives of the defining equation vanish at this point.
Zero conditions for functions are always closed conditions.
Hence non-singularity is an open condition
on the individual curve. In fact, it is even an open condition
with respect to the variation of the point on the base scheme.
The curve $X^2-Y^2-5=0$ is a non-singular curve over $\Q$.
The openness principle applied to the base scheme says that
there are only finitely many primes for which the fibre will
become singular.
Here it is quite easy to  calculate these primes. Let $\ f(X,Y)=X^2-Y^2-5\ $
be the defining equation. Then $\frac {\partial f}{\partial X}=2X$
and $\frac {\partial f}{\partial Y}=2Y$. For $p=2$ both partial
derivatives vanish at every point on the curve (the fibre). Hence every point
of the fibre is a singular point. This says that   the fibre over the
point $[(2)]$ is
a multiple fibre. In this case we see immediately
$(X^2-Y^2-5)\equiv (X+Y+1)^2\mod 2$. This special fibre is
$\ \Spec(\F_2[X,Y]/((X+Y+1)^2)\ $ which is a non-reduced scheme.
For $p\ne 2$ the only candidate for a singular point is $(0,0)$.
But this candidate  lies  on the
curve if and only if
$5\equiv 0 \mod p$ hence only for $p=5$. In this case we get one
singularity. Here we calculate
that $\ (X^2-Y^2-5)=(X+Y)(X-Y)\mod 5$. Altogether we obtain that nearly
every fibre is a non-singular conic. Only the fibre over $[(2)]$ is a
double line and the fibre over $[(5)]$ is a union of
two lines which meet at one
point.
\subhead
4. Other objects
\endsubhead
In lecture 5 we already said that moduli problems
(degenerations etc.) can be  conveniently be described as functors.
It is not always possible to find a scheme representing a certain
moduli functor.
To obtain a representing geometric object it is sometimes necessary to
enlarge the category of schemes by introducing more general objects
like algebraic spaces and algebraic stacks.
It is quite impossible even to give the basics of their definitions.
Here let me only say that in a first step it is necessary to introduce
a finer topology on the schemes, the {\sl etale} topology.
With respect to the etale topology one has more open sets.
Schemes are ``glued'' together from affine schemes
using  algebraic morphisms. Algebraic spaces are
objects where the ``glueing maps'' are more general maps
(etale maps).
Algebraic stacks are even more general than algebraic spaces.
The typical situation where they occur is in connection with
moduli functors. Here one has a scheme which represents a set
of  certain objects.
If one wants to have only one copy for each isomorphy class of the objects one
usually has to divide out a group action.
But not every orbit space of a
scheme by a group action can be made to a scheme again.
Hence we indeed get new objects.
This new objects are the algebraic stacks.
\nl
Let me here only give a few references.
More information on algebraic spaces you can find in the book of
Artin \cite{\ARAS} or Knutson \cite{\KNAS}.
 For stacks  the appendix of  \cite{\VIST} gives a very short
introduction and some examples.
%
%
%
\vskip 1cm
\head
{\bf References}
\endhead
\vskip 0.7cm

\refARAS

\refART

\refBOGARE

\refEHS

\refGOWA

\refGHPA

\refEGAI
\np
\refEGAA

\refHAG

\refKNAS

\refKUKA

\refMAQG

\refMANG

\refMADR

\refMRB

\refMPMP

\refROS

\refSCHLRS

\refVIST

\refWEQG
\enddocument
\bye